\documentclass[acmsmall, screen]{acmart}

\AtBeginDocument{%
  }

\setcopyright{acmlicensed}
\copyrightyear{}
\acmYear{}
\acmDOI{XXXXXXX.XXXXXXX}

\acmJournal{JACM}
\acmVolume{37}
\acmNumber{4}
\acmArticle{111}
\acmMonth{8}

\usepackage{listings}
\usepackage{xcolor}
\usepackage[ruled,vlined,linesnumbered,commentsnumbered]{algorithm2e}
\usepackage{amsthm}
\usepackage{amsmath}
\usepackage{multirow}
\usepackage[]{hyperref}
\usepackage{multirow}
\usepackage{wrapfig}
\usepackage[most]{tcolorbox}

\theoremstyle{definition}

\theoremstyle{remark}

\definecolor{codegreen}{rgb}{0,0.6,0}
\definecolor{codegray}{rgb}{0.5,0.5,0.5}
\definecolor{codepurple}{rgb}{0.58,0,0.82}
\definecolor{backcolour}{rgb}{0.94,0.94,0.94}
\definecolor{codeblue}{rgb}{0,0,1}
\definecolor{codegold}{rgb}{0.85,0.65,0.13}

\lstdefinestyle{mystyle}{
    backgroundcolor=\color{backcolour},   
    commentstyle=\color{codegreen},
    keywordstyle=\color{magenta},
    numberstyle=\tiny\color{codegray},
    stringstyle=\color{codepurple},
    basicstyle=\ttfamily\footnotesize,
    alsoletter={_},                 
    emphstyle=[2]\color{codeblue},   
    emphstyle=[3]\color{codegold},   
    emphstyle=[4]\color{magenta},
    breakatwhitespace=false,         
    breaklines=true,                 
    captionpos=b,                    
    keepspaces=true,                 
    numbers=left,                    
    numbersep=5pt,                  
    showspaces=false,                
    showstringspaces=false,
    showtabs=false,                  
    tabsize=2
}

\lstdefinelanguage{Coq}{
    keywords={Inductive, Definition, Fixpoint, Lemma, Theorem, Proof, Qed, intros, apply, exact, reflexivity, simpl, rewrite, assumption, constructor, exists, forall, Prop, Type, Set, fun, match, with, end, Require, Import},
    keywordstyle=\color{blue}\bfseries,
    identifierstyle=\color{black},
    comment=[s]{(*}{*)},            
    commentstyle=\color{green!50!black}\ttfamily,  
    stringstyle=\color{red}\ttfamily,
    basicstyle=\small\ttfamily,
    showstringspaces=false
}

\author{Bodhisatwa Chatterjee}
\orcid{https://orcid.org/0000-0003-3098-6256}
\email{bodhi@gatech.edu@}
\affiliation{%
  \institution{Georgia Institute of Technology}
  \city{Atlanta}
  \country{USA}
}

\author{Drew Zagieboylo}
\orcid{}
\email{dzagieboylo@nvidia.com}
\affiliation{%
  \institution{NVIDIA Research}
  \city{Westford}
  \country{USA}
}

\author{Sana Damani}
\orcid{}
\email{sdamani@nvidia.com}
\affiliation{%
  \institution{NVIDIA Research}
  \city{Santa Clara}
  \country{USA}
}

\author{Siva Hari}
\orcid{}
\email{shari@nvidia.com}
\affiliation{%
  \institution{NVIDIA Research}
  \city{Santa Clara}
  \country{USA}
}

\author{Christos Kozyrakis}
\orcid{https://orcid.org/0000-0002-3154-7530}
\email{christos@cs.stanford.edu}
\affiliation{%
  \institution{Stanford University \& NVIDIA Research}
  \city{Santa Clara}
  \country{USA}
}

\begin{document}

\title[ProofWright: Towards Agentic Formal Verification of CUDA]{ProofWright: Towards Agentic Formal Verification of CUDA}



\begin{abstract}
Large Language Models (LLMs) are increasingly being used to automatically generate optimized CUDA kernels, substantially improving developer productivity. However, despite rapid generation, these kernels often contain subtle correctness bugs and lack formal safety guarantees. In practice, runtime testing remains the primary validation approach, but it provides limited input coverage and is vulnerable to reward hacking, which can mask incorrect behavior. While formal verification tools offer an alternative with greater correctness guarantees, they often rely on manual annotation 
that unfortunately cannot keep pace with rate of LLM-generated codes, creating a critical validation bottleneck.

To address these challenges, we present \textit{ProofWright}, an automated, feedback-driven verification methodology which uses LLMs for automatic formal verification of CUDA kernels. While naive prompt engineering techniques fail at the task of automatic formal verification, ProofWright incorporates experiential learning, static analysis and program translation techniques with SMT solvers and theorem provers to provide memory safety, thread safety, and semantic correctness guarantees for LLM-generated CUDA kernels automatically and with low overheads. On KernelBench L1, ProofWright verifies memory safety and data race freedom for 74\% of the kernels generated by an LLM, uncovers subtle correctness errors missed by conventional testing, and establishes semantic equivalence for a subclass of element-wise kernels. With a modest average overhead of $\sim$3 minutes per kernel, ProofWright demonstrates that scalable, automated formal verification of LLM-generated GPU code is feasible and offers a path toward trustworthy high-performance code generation without sacrificing developer productivity. 

\end{abstract}

\maketitle
 
\section{Introduction}
\label{sec:intro}

Large-Language Models (LLMs) have been remarkably successful in a wide range of domains spanning from natural language processing to software engineering. For the past couple of years, agent-based AI systems have been extensively used for generating and optimizing programs \cite{cummins2025llm, taneja2025llm, huang2024effilearner, peng2024perfcodegen}. This includes agentic-AI systems for GPU program optimizations \cite{damani2024warpdrive, lange2025aicuda, baronio2025kevin32b}, where the goal is to automate the tedious and difficult process of generating highly efficient GPU kernels by leveraging LLMs. These agentic systems follow an iterative approach of code generation: the first step involves generating GPU kernel(s) based on a given user specification or prompt. The generated code is then subjected to validation and performance testing; based on the testing feedback, the kernels can then be iteratively refined until the feedback is satisfactory.

However, owing to the lack of correctness guarantees in the generated code, gaining explicit trust in such agentic systems remains an open challenge. This issue is particularly pronounced for highly optimized GPU kernels, where aggressive performance tuning often introduces subtle synchronization and memory-access errors that may not be caught in testing. Such kernels can be inherently complex as they often combine massive parallelism with strict synchronization requirements, intricate memory hierarchies, and numerous low-level optimizations. As a result, validating the correctness of GPU programs raises deeper concerns about how much confidence can be placed in the testing and verification performed by agentic systems.

Current agentic-AI systems \cite{lange2025aicuda, baronio2025kevin32b, cummins2025llm} reason about the correctness of the generated code by performing compilation checks and functional unit tests. This involves compiling the generated code, executing the program with predefined inputs, and then numerically comparing the program’s output against a reference implementation. In case of failures, the error feedback is used by the agent to modify the kernel and hopefully resolve the bug causing the error. While this approach can increase code quality, it often fails to fully validate all kernels. Functional correctness on a test suite is not sufficient to rule out rare but potentially dangerous behaviors such as data races and out-of-bounds memory accesses (see Listing \ref{code1}) and the complexity of modern GPU kernel optimizations make writing high-coverage test suites even more challenging. This can be further exacerbated by poorly constructed testing environments that allow the agent to \textit{reward hack}: a case where the agent produces an output that \textit{only} passes the validation tests but otherwise does not fit the intended specifications. Some common examples of reward hacking include copying the output of the reference implementation, or removing necessary synchronization primitives to improve performance metrics.

Although some reward hacking (such as the trivial solution-copying example in Appendix D) can be easily fixed, guaranteeing the absence of subtle implementation bugs is far more challenging. Conventional testing strategies such as \textit{input-based dynamic testing} \cite{NVIDIA_Compute_Sanitizer, peng2018curd, kamath2020scord, kamath2021iguard, tarek2023cucatch, jacobson2024hirace} and \textit{symbolic execution} \cite{Li:2012:GKV:2145816.2145823, Li2012Parameterized, Li2010ScalableSMT} are capable of detecting memory safety violations and/or data races and thus improve on traditional functional testing. However, these are both more expensive to execute than typical tests and also still lack completeness guarantees; they cannot examine all possible program paths or thread interleavings (\S \ref{subsec:currtest}). To adopt AI-generated GPU kernels in high-integrity applications (such as automotive and avionics systems~\cite{aspetakis2024}), we must utilize tools that can formally rule out unsafe behaviors and demonstrate strong correctness guarantees.


To this end, we propose a new validation strategy for LLM-generated CUDA codes that complements traditional testing approaches with formal verification techniques to provide functionality and safety guarantees. In general, formal reasoning of GPU programs is particularly challenging due to concurrent execution of thousands of parallel threads, tight coupling of the programming model with various memory spaces, and varied synchronization primitives and patterns. Existing work on GPU code verification \cite{Li2010ScalableSMT, Betts2012GPUVerify, Li2012Parameterized, Li:2012:GKV:2145816.2145823, Berkovich2013GPUbased, ZhengRLDS15, Ferrell2019CUDA} either focus on verifying a small subset of properties such as \textit{bank conflicts}, \textit{improper synchronizations}, or rely on the user to provide hints for establishing a limited set of properties. Identifying the right tools for specific use cases, or constructing verifiable annotations can be tedious, error-prone, and requires expert knowledge. Furthermore, the demand for rigorous verification of GPU programs has grown substantially with the advent of agentic-AI systems, which can generate kernels at an unprecedented scale that far outstrips the ability of human experts to formally verify them. With future LLMs increasingly trained on these AI-generated codes \cite{lange2025aicuda}, ensuring correctness and reliability is  paramount. 

Recent works \cite{tu2025agentic, kumar2025saarthi, pothireddypalli2026agentic} have explored the use of agentic systems for formal verification, where LLMs are used to automatically orchestrate verification tools; however, our experiments (\S \ref{RQ3}) show that \textit{naively integrating formal tools with LLMs via prompt engineering alone fails spectacularly}. To address this challenge, we present \textbf{ProofWright}, which is an  automated, feedback-driven verification methodology built on new analysis tools, libraries and agents. It formally reasons about LLM-generated CUDA codes by automatically establishing two key program properties: \textit{implementation correctness} and \textit{semantic equivalence}. The implementation correctness of the generated GPU code is ascertained by proving \textit{thread safety} (i.e., the absence of data races) and \textit{memory safety} (i.e., the absence of illegal memory accesses), while the semantic equivalence is established by checking that the generated code adheres to its original functional specification. 

Our core technical contributions \footnote{ProofWright will be open-sourced and submitted for artifact evaluation} include:
\begin{itemize}
    \item An Agentic Formal Verification Framework that uses in-context learning and past experience to automatically generate annotations for an SMT-solver based formal verification tool (VerCors) to prove memory safety and data-race freedom for CUDA kernels.
    \item An Agentic Semantic Equivalence Framework that uses static analysis techniques to extract program specifications, a novel mathematical abstractions library containing formal implementations of tensor operations, and a theorem prover (Rocq) to automatically check if the generated CUDA kernel is semantically equivalent to the provided specification.
\end{itemize}

We evaluated ProofWright on the \textit{KernelBench} benchmark \cite{ouyang2025kernelbenchllmswriteefficient}, which provides a comprehensive suite of PyTorch programs designed to evaluate the ability of LLMs to generate equivalent CUDA kernels that are both performant and correct. Our experiments show that ProofWright successfully establishes memory and thread-safety guarantees for up to \textit{74\%} of KernelBench L1 programs and can ascertain semantic equivalence in \textit{14\%} of KernelBench L1 programs.


\section{Background and Motivation}
\label{sec:back}

This section gives a brief description of agent-based systems for GPU code generation, and motivates the shortcomings of their testing methodologies. We then describe the rationale for employing an agentic approach for verification, and introduce the existing tools used in this paper.  
\subsection{Agentic Systems for GPU Code Generation \& Optimization}
Agentic systems for GPU code generation leverage LLMs to automatically produced highly optimized GPU kernels from high-level specifications. The input specification may be unstructured natural language, but we focus on systems that take a structured high-level program as the desired output specification. PyTorch~\cite{2019pytorch} is the most popular language for this use case due to its prevalent usage in GPU-accelerated computation for AI; it used as the specification and reference implementation in the KernelBench benchmark suite~\cite{ouyang2025kernelbenchllmswriteefficient} that is targeted by several existing agentic systems~\cite{damani2024warpdrive, lange2025aicuda} for CUDA code generation.

Once provided the specification (which is often augmented with a system prompt and other contextual information), the agentic systems use LLMs to generate candidate GPU kernels that are then iteratively tested and refined based on feedback from various tools such as static analyzers, compilers, functional tests, performance profilers, and even other LLM judges~\cite{damani2024warpdrive, lange2025aicuda}. Some agentic systems structure the workflow and allow AI models to make limited (or no) judgments of which tools to use and when; others allow the LLM to plan its own execution and freely use tools and informational resources at its own discretion. In all cases, some level of functional testing is typically deterministically invoked on the final result; if it passes the tests it is accepted, otherwise the system either retries or returns no result.


\subsection{Shortcomings of Current Testing Methodologies: False Negatives, Code Coverage, Compatibility Issues}
\label{subsec:currtest}

CUDA programs are prone to a wide range of errors due to their massively parallel nature and the complex architecture of modern GPGPUs. The CUDA programming model exposes different memory spaces (global, shared, local) with distinct address spaces and scopes, which can potentially lead to subtle bugs that are often difficult to detect and reproduce. Furthermore, CUDA programs freely manipulate data with raw pointers which comes with all of the well known pitfalls of other memory unsafe languages. To detect these issues, programmers often use testing and debugging tools that can be broadly divided into two classes:

$\blacksquare$ \textbf{Input-based Dynamic Testing Tools}:  These tools detect concurrency bugs by executing GPU programs with concrete inputs, and dynamically monitoring inter-thread interactions to expose data races and shared-memory conflicts across thread execution schedules. This includes widely-used commercial tools such as NVIDIA's \textit{Compute Sanitizer} \cite{NVIDIA_Compute_Sanitizer}, and popular academic tools \cite{peng2018curd, kamath2020scord, kamath2021iguard} for detecting data-races. However, these tools can only detect race conditions and illegal memory accesses for a given input. For instance, the LLM-generated kernel depicted in Listing \ref{code1} attempts to optimize the sigmoid computation by vectorization, where each thread processes 4 consecutive FP16 elements. However, the tail-processing logic (line 6) incorrectly assigns the first thread ($threadIdx.x = 0$) of multiple (remainder) thread blocks to perform the write operation on $y[tail]$ (line 10), leading to a race-condition. This condition will only arise if the original input length $N$ is not a multiple of 4, and will elude the input-based dynamic testing tools for other inputs, resulting in a false negative \footnote{We were unable to detect this error using NVIDIA compute-sanitizer's racecheck tool}.

\begin{lstlisting}[language=C++,caption={Sigmoid Kernel with race condition generated by GPT4.1, where the tail processing logic incorrectly distributes work across blocks, causing threads from blocks 0, 1, and 2 to collide on the write operation. This passes both compilation and muliple trials of unit tests.},label=code1,frame=tlrb, numbers=none]
__global__ void sigmoid_kernel(..) {
    int idx = blockIdx.x * blockDim.x * 4  + threadIdx.x * 4; 
    int64_t N4 = N & (~int64_t(3)); 
    if (idx < N4) {..}
    // Processing Tail Elements (< 4) 
    if (threadIdx.x == 0) {
        int idx2 = blockIdx.x * blockDim.x + threadIdx.x;
        for (int tail = N4 + idx2; tail < N; ++tail) { 
            float xf = __h2f(x[tail]); 
            y[tail] = __f2h(sig(xf));
}}}
\end{lstlisting}
$\blacksquare$ \textbf{Symbolic Execution-based Tools}: These tools systematically explore the program paths by representing the program inputs as symbolic variables, constructing logical constraints that describe all feasible execution behaviors, and using SMT solvers to reason about correctness across both CPU and GPU code. However, such tools only support simple atomic operations, and do not scale beyond smaller kernels. 

Furthermore, none of these aforementioned techniques can actually determine whether the generated code semantically satisfies its intended specification. For instance, the LLM-generated kernel shown in Listing 3 (Appendix D) was supposed to implement a vector-add operation, as specified by the user. As illustrated, the kernel does not implement the intended computation, and instead generates a `hacked kernel' which passes both unit and compilation tests. Moreover, existing testing strategies focus primarily on detecting the presence of errors rather than explaining their cause or providing actionable feedback. This limitation makes them inadequate in an agentic context, where the feedback loop back to the agent is necessary for iterative code refinement.

\subsection{VerCors Program Verifier}
\label{subsec:vercors}

VerCors \cite{blom2014vercors, armborst2024vercors} is a deductive program verifier for concurrent code. Similar to popular languages such as Dafny~\cite{leino2010dafny}, VerCors can prove that programs adhere to Hoare-style contracts; unlike other verification languages, VerCors can also prove memory safety and data race freedom of programs written in unsafe languages such as C and CUDA-C. VerCors provides front-ends for a variety of languages (including CUDA) which it translates to an intermediate verification language.

VerCors enables fine-grained reasoning about concurrent access to shared resources via permission-based \textit{concurrent separation logic}\cite{haack2015permission}, where a permission of 1 enables write access and any non-zero fraction allows only read permission. VerCors guarantees race freedom by maintaining the invariant that the total permission for each memory location never exceeds 1. 

To make verification possible, programmers must add annotations that guide how to split permissions across threads when they are launched or at synchronization barriers. As part of a CUDA kernel contract, the pre-condition is quantified over all threads that are launched, and must define how the initial permissions are split between threads. For instance, the following pre-condition requires that each thread has write access to a location in the array \texttt{C} specified by its canonical CUDA thread id:
\begin{lstlisting}[language=C++, numbers=none]
//@ requires Perm( C[blockDim.x*blockIdx.x + threadIdx.x], write );
\end{lstlisting}
Notably, VerCors must prove that the memory locations accessed by difference threads are disjoint (since they each require write permission) and in-bounds. For more complex indexing patterns, this is undecidable in general and users must specify hints to the verifier via helper functions or frame statements to make verification more likely to terminate. Listing \ref{lst:vc1} demonstrates an example of using these features in a toy kernel. However, even with these hints, it is not always possible to guide the verifier into success and some correct programs may not be verifiable.

\begin{lstlisting}[language=C++,caption={An abbreviated example of how to specify write permissions with complex indexing patterns in CUDA-VerCors. The contract of the acc2d function guides VerCors with relevant facts and the \{: :\} syntax denotes triggers that indicate to VerCors when to instantiate hidden quantifiers during verification. CUDA launch parameters and pointer validitity requirements are excluded for brevity.},label=lst:vc1,frame=tlrb,numbers=none,float=htpb]
//@ requires 0 <= row && row < sz && 0 <= col && col < sz;
//... other parts of contract omitted
//@ ensures 0 <= result && result < sz * sz;
/*@ pure @*/ int acc2d(int row, int col, int sz) { return row * sz + col; }

//other parts of the contract omitted...
//@ requires (\forall* int i; 0<=i && i<sz; Perm(A[i], read) ** Perm(B[i], read));
// \gtid(n) is shorthand for the CUDA id expression in a specific dimension
//@ requires Perm( {: C[acc2d(\gtid(0), \gtid(1), sz)] :} , write);  
__global__ void toy_kernel(int* A, int* B, int* C, int sz){
    int row = blockDim.x * blockIdx.x + threadIdx.x;
    int col = blockDim.y * blockIdx.y + threadIdx.y;
    C[acc2d(row, col, sz)] = A[row] + B[col];
}}}
\end{lstlisting}

\subsection{Rocq Theorem Prover}
The \textit{Rocq Prover}\footnote{formerly known as Coq}~\cite{the_coq_development_team_2024_11551307} is an interactive theorem-proving environment that allows users to formally specify software properties as mathematical theorems and lemmas, and develop \textit{machine-checkable} proofs through an extensible system of tactics. It is a \textit{dependently typed} programming framework, which means that the types can depend on values (such as a vector type $A$ parameterized by its length $n$: \texttt{Vector A n}) which allows programs and proofs to be expressed within the same formal system. At its core, the Rocq Prover uses Gallina, a high-level specification language that provides the syntax and semantics for defining functions, data types, and logical propositions.

\section{ProofWright Framework: An Overview} 

\begin{figure} \centering\includegraphics[width=1.0\textwidth]{figs/proofwright.png}
\caption{\textit{ProofWright as Part of an Agentic Code Generation Workflow}. ProofWright integrates into an agentic code generation workflow, where an AI agent produces optimized CUDA kernels from PyTorch code and verifies them via semantic equivalence (Rocq) and memory/thread safety (Vercors). This feedback loop ensures correctness and safety before final deployment.}
\label{fig:overview}
\end{figure}

\autoref{fig:overview} shows how our framework fits within fits within the overall agentic code generation workflow. We develop agentic flows that formally reason about LLM-generated CUDA codes by automatically establishing three key program properties: memory safety, thread safety and semantic equivalence. Our framework consists of two major components described below. 

\textbf{Agentic Formal Verification Framework} (\S \ref{sec:vercorsagent}): We build an agentic flow that automatically utilizes a deductive program verifier to establish safety properties of a given CUDA code. The agent is responsible for producing program annotations that specify the expected program behavior for memory and thread safety and that assist the verifier via hints. In this work, we leverage the \textit{VerCors verifier}, which is a SMT-based tool for deductive verification of concurrent and parallel software (\S \ref{subsec:vercors}). The VerCors agent consists of three key components: 1) a \textit{Knowledge Base}, consisting of annotation syntax and rules obtained from VerCors documentation, 2) the \textit{Annotation Guide}, an automatically generated document that summarizes a general verification procedure and dynamically updated to incorporate newly discovered concepts after successful verification, and 3) a minimal database of known errors and their corresponding fixes. The \textit{VerCors Agent} shows strong learning capability to automatically generalize across experiences and generate memory and thread safety annotations  without any manual intervention. 

\textbf{Agentic Semantic Equivalence Framework} (\S \ref{sec:roqagent}):  An agentic framework that automatically constructs proofs for determining whether AI-generated CUDA code functionally adheres to a given user specification. This framework has three major components: (1) a \textit{static analysis tool} (\S \ref{subsec:pytorchtool}) that abstracts the initial input program into a computation graph to capture its intended functional specification, (2) an \textit{operations mapping tool} (\S \ref{subsec:rocqtrans}) which translates the specifications graph into mathematical properties, and (3) the \textit{Rocq Agent} (\S \ref{subsec:rocqlowering}), which generates alternative theorems and properties corresponding to the given specifications and automatically constructs proofs to establish equivalence between them. The verified theorems are then lowered into program annotations and validated by a formal verifier. The \textit{Rocq agent} utilizes the \textit{Rocq Theorem Prover}, an interactive proof assistant with a dependently-typed language. The \textit{Rocq Agent} is able to generate a group of Rocq tactics in order to construct the theorem proofs. 

\textit{Deployment}: In a typical GPU agentic code generation pipeline, numerous variants of the kernel may be explored at each iteration \cite{lange2025aicuda}; in such a scenario it might be practical to use ProofWright for ascertaining formal guarantees after every few iterations. Ideally, in an end-to-end workflow, we envision that ProofWright could be used in the later-stages of the codegen as a high-guarantee filter to promising candidates, which have already passed the faster, lower assurance conventional testing such as compilation checks and unit tests.


Our agentic solution is reliable because we use formal verification tools and mathematical proofs. If any of our agents hallucinate and generate incorrect annotations or proofs, they will simply fail to verify. That is, \textit{we will never verify a program that is incorrect.} The following sections describe each component in detail.

\section{Agentic Formal Verification Framework: Establishing Thread \& Memory Safety Guarantees}
\label{sec:vercorsagent}

\begin{figure*}[htbp]
\centering\includegraphics[width=1.0\textwidth]{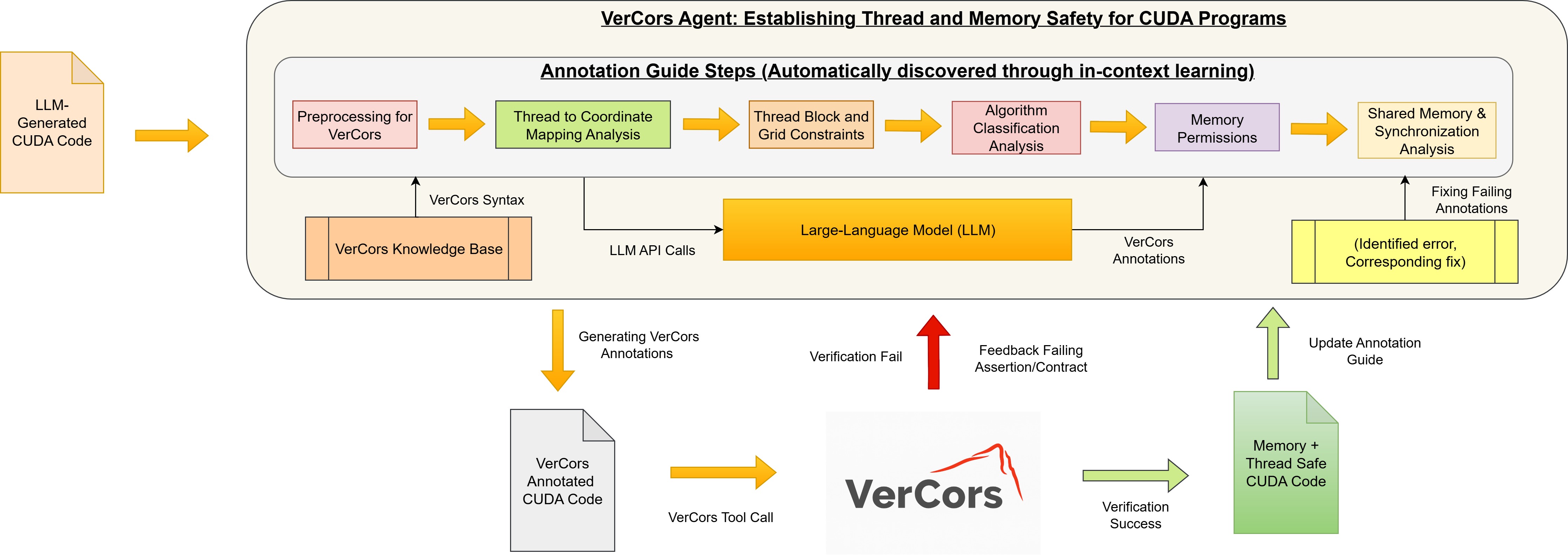}
\caption{\textbf{\textit{Agentic Formal Verification Framework} for establishing memory and thread safety for CUDA Programs}. The LLM-generated CUDA kernels are first minimally preprocessed, and then the safety annotations are generated in a step-by-step fashion by the Vercors agent, according to the automatically crafted \textit{annotation guide}. The annotated program is then verified by calling the VerCors tool.}
\label{fig:vercors1}
\end{figure*}

This section introduces the \textit{VerCors Agent}, an LLM-based solution that automatically inserts Vercors annotations in CUDA programs for formal verification. Figure \ref{fig:vercors1} shows the high-level agentic workflow. The VerCors Agent is an iterative LLM-based agent with access to a static, hand-written \textit{Knowledgebase}, a dynamic, LLM-generated \textit{Annotation Guide} that learns verification strategies from experience, and the VerCors program verifier which provides feedback to the agent.
The following subsections describe the agent’s internal architecture and verification methodology in detail.

\subsection{Verification Procedure}

As discussed in Section~\ref{subsec:vercors}, verifying CUDA programs with VerCors requires the agent to generate formal contracts for each target kernel, specifying the preconditions necessary for safe invocation and the post-conditions guaranteed upon successful execution. For the purposes of memory and thread safety, we do not need the contracts to specify any meaningful postconditions; VerCors inherently ensures that these safety properties hold throughout the program’s execution as long as the preconditions have been satisfied. Since it is not generally possible to differentiate between a program that does not actually satisfy its contract and an instance of imprecision in the verification process, we ask the agent to essentially infer the \emph{weakest pre-conditions} possible such that the kernel contract is satisfiable. This approach enables the following workflow:

\begin{itemize}
    \item Use the VerCors Agent to attempt to infer the weakest pre-conditions necessary to satisfy thread and memory safety.
    \item If accepted by VerCors, any invocation of the kernel that satisfies the precondition is safe; this can also be checked by VerCors as long as a sample program is supplied.
    \item If not, we conservatively assume that it is \emph{never} safe to invoke the kernel.
\end{itemize}
Note that we do not confirm that the kernel contracts are indeed \emph{weakest} preconditions; however, in our evaluation (\S\ref{sec:eval}) we do execute basic checks to ensure that the contracts produced are not trivially unsatisfiable.

\paragraph{An Example Kernel Verification Walkthrough}

We direct the VerCors agent to jointly infer these contracts and generate any other annotations in the kernel body needed to verify the kernel. To give the reader intuition for the knowledge needed by VerCors agent to complete this task, we describe a representative kernel and its associated VerCors annotations in Listing~\ref{vcexample}.

\begin{lstlisting}[language=C++,frame=tlrb, caption = {Example contract and annotations generated by VerCors Agent to establish memory and thread safety. The annotations before the function header define the contract and the VerCors comments in the body serve as hints to aid the automated verifier.},label=vcexample,escapechar=|,float=htpb]
/*@
context_everywhere (\forall* int i; 0 <= i && i < N*N; Perm({:A[i]:}, read)); | \label{line:reada} |
context_everywhere (\forall* int i; 0 <= i && i < N*N; Perm({:B[i]:}, read)); | \label{line:readb} |
context_everywhere (access_func(..) < N && access_func(..) < N) ==> 
    Perm({: C[access_func(access_func(..), access_func(..), N, N)] :}, write); | \label{line:writec} |
@*/
__global__ void square_matmul_basic_kernel(float* A, float* B, float* C, int N) {
    int row = blockIdx.y * blockDim.y + threadIdx.y;
    int col = blockIdx.x * blockDim.x + threadIdx.x;
    if (row < N && col < N) { | \label{line:if} |
        float sum = 0.0f;        
        //@ loop_invariant 0 <= k && k <= N;
        for (int k = 0; k < N; k++) {
            //@ assert access_func(row, k, N, N) == row * N + k;
            //@ assert access_func(k, col, N, N) == k * N + col;
            sum += A[row * N + k] * B[k * N + col];
        }
        //@ assert access_func(row, col, N, N) == row * N + col; | \label{line:assert} |
        C[row * N + col] = sum; | \label{line:execcwrite} |
    }
}
\end{lstlisting}

The agent needs to (1) identify all memory operations in the kernel, 
(2) determine whether each access is a read or write, and then (3) describe the memory location(s) based on the thread and block identifiers and conditions under which the memory is accessed. For example, many different locations in arrays \texttt{A} and {B} are read by each thread, so in lines~\ref{line:reada} and ~\ref{line:readb} the agent conservatively requires that every element in the bounds of those arrays are readable by every thread.
In this case the \texttt{context\_everywhere} annotation is used to specify that this is not only a precondition, but also a postcondition and loop invariant.

However, the agent cannot make the same requirement for the array \texttt{C} since write permission is needed to execute the assignment on line~\ref{line:execcwrite}; in this case the agent adds a \emph{conditional} precondition on line~\ref{line:writec}. The condition matches the if statement on line~\ref{line:if} and the permission location matches the index dereferenced on line~\ref{line:execcwrite}. Note that a helper function \texttt{access\_func} (which is defined by the agent elsewhere and \texttt{ensures} that the index is in-bounds) is used to help guide VerCors' solver on line~\ref{line:assert}.
Although omitted from Listing~\ref{vcexample}, the agent also needs to infer minimum necessary array sizes, constraints on the CUDA block and grid dimensions, and reasonable bounds on any other function parameters. Furthermore, for examples with synchronization, the agent must specify how permissions change over time at each thread barrier.

The remainder of this section describes both the challenges in building the VerCors agent and how we provide it with the information necessary to achieve its goal.

\subsection{Tackling Low-Resource VerCors Syntax}

We found that most foundation models could not reliably produce syntactically valid VerCors-annotated CUDA.
This is unsurprising as there is only a small corpus of publicly available VerCors code, even compared to other program verification languages like Dafny~\cite{leino2010dafny} or F*~\cite{swamy2016fstar}, and it is unlikely that any LLMs have been trained sufficiently on VerCors source code. Hence, we employ in-context learning to improve the model's ability to generate VerCors annotations.

\subsubsection{VerCors Knowledge Base: In-Context Learning}

As a first step, we build a VerCors knowledge base using traditional prompt engineering techniques
which includes (1) context: a stripped-down version of VerCors' official documentation, (2) few-shot examples: hand-annotated, verified CUDA kernels, and (3) a set of error-fix pairs that teach the agent how to handle VerCors errors.

The VerCors knowledge base includes examples about the general VerCors syntax as well as GPU-specific examples. Appendix E.1 contains two listings which show how permissions syntax are formed and evolved over constructs such as barrier synchronizations, etc.
Next, if the \textit{VerCors Agent} fails to generate verifiable annotations on the first attempt, it must perform \textit{proof repair}, refining its annotations iteratively based on the feedback received from the verifier. 
Although errors returned from the verifier were sometimes immediately helpful (e.g., parsing errors), often others were more ambiguous (e.g., ``Insufficient permission to assign to field") and the agent struggled to correct them. Therefore, we provided a small set of failed verification attempts, the corresponding error message, and a fixed version of the code along with a textual explanation of the fix as part of the VerCors knowledge base. \autoref{lst:errorexample} describes the layout of entries in this dataset. 

The \texttt{ORIGINAL} tag indicates the code that caused the error which failed verification. The \texttt{ERROR} tag indicates the error received from VerCors verbatim; these errors include a stack-trace like explanation of which proof obligations failed and the related CUDA source code. The \texttt{EXPLANATION} is a human-written explanation of (1) how the error message relates to the existing code and (2) an explanation of what the most reasonable fix is for the error. Lastly the \texttt{CODE} tag precedes a fixed version of the code that incorporates the changes proposed in the explanation.

\begin{figure*}[h]
    \begin{lstlisting}[keywords={ORIGINAL,ERROR,CODE,EXPLANATION},frame=tlrb,numbers=none]
    #ORIGINAL:
    <The kernel with VerCors annotations that caused the error>
    #ERROR:
    <The VerCors error message and associated source information>
    #EXPLANATION:
    <An explanation of both why the error occurred and how best to fix it.
    #CODE
    <The code with the described modification>\end{lstlisting}
    \vspace{-1em}
    \caption{The structure of verification error repair examples provided as part of the VerCors knowledge base.}
    \label{lst:errorexample}
\end{figure*}


We generated this dataset by recording the manual verification process for a complex kernel that required many rounds of iteration. While a verification expert could likely have verified this with few iterations, we intentionally started with a basic contract and corrected each error that we received one-at-a-time. This ensured that each example in the dataset was focused on a singular error, and enabled us to elicit a reasonable set of examples while analyzing few kernels. The entire knowledge base is provided to the agent as part of its prompt. It consists of $\sim$ 12K words, a relatively small fraction of the approximately 300K word VerCors wiki. 

\subsubsection{VerCors Annotation Guide: Learning from Experience}

While the static VerCors knowledge base described in the previous section provides a starting point to the VerCors agent, it has several limitations. First, it provides syntax and examples and not a methodology on how to approach verification of a new program. Due to the lack of exemplar data, we found this resulted in pattern-matching to generate annotations which does not scale well to new kernels that may have different control flow, data mapping, or synchronization patterns. Second, even when the agent successfully generated annotations after several iterations of the feedback loop, it fails to learn from experience and may repeat similar errors on future verification problems.

To overcome these limitations, we supplement the agent with a dynamic \textit{VerCors Annotation Guide} where the agent automatically distills knowledge based on human-provided few-shot examples as well as past experience into a generalized verification recipe. It is a LLM-generated, dynamically maintained document that can be updated after successful verification attempts to incorporate newly discovered concepts and verification patterns. A detailed description of how the annotation guide was constructed is presented in the evaluation section (\S \ref{sec:eval}). This Annotation Guide serves as the agent’s core reasoning framework and helps it successfully generate VerCors-compatible specifications across a diverse set of CUDA kernel structures. Further investigation into the generated guide showed that the agent learned some interesting core verification principles across multiple sessions. We now briefly describe some of these insights.



$\blacksquare$ \textit{Algorithm Class-Aware Verification Scheme}: The agent discovered that different algorithm families exhibit distinct memory access patterns and synchronization primitives, requiring specialized verification strategies beyond generic templates. To address this, the agent classifies kernels into different families such as element-wise, convolution-like, reduction, etc. Each family follows unique permission and indexing schemes—ranging from simple one-to-one thread-to-element mappings to complex shared-memory coordination in convolution operations. In some cases, the agent even employs a hybrid classification strategy, combining features from multiple families to accurately capture mixed computational behaviors within a single kernel.

$\blacksquare$ \textit{Thread-to-Data Mapping}: The agent observed that the verification strategy and corresponding annotations vary significantly with the thread-data mapping pattern used in CUDA kernels. Different index-mapping schemes require distinct reasoning regarding memory accesses, which often necessitates specialized \textit{helper functions} to express the mapping between thread and data coordinates and aid in bounds checking. These functions help with SMT instability as they define postconditions that can serve as hints to the verifier and its underlying solver. While some of the few-shot examples did include use of similar functions, we did not explicitly instruct the agent when or how to use them; it decided that they were helpful and proposed adding them to the annotation guide on its own. Listing~\ref{code4} illustrates how the agent generates specialized helper functions for a batched 4D array access pattern, which was not demonstrated in the few-shot examples.


$\blacksquare$ \textit{Determining Thread Block \& Grid Constraints}: The agent identifies thread block and grid dimension configurations based on the kernel’s computational structure (e.g., element-wise, matrix multiplication, or reduction operations). These annotations often fix certain dimensions to constant values (e.g., $blockDim.x == 256$, $gridDim.z == 1$) to simplify the verification process, while still aiming to maximize problem-space coverage by generalizing over unconstrained dimensions. This balance between fixed and parameterized dimensions enables both tractable and broadly applicable verification conditions. Further, the agent includes annotations to ensure that all data pointers are non-null, array dimensions are positive, and memory allocations satisfy the length constraints. 

\begin{lstlisting}[language=C++,caption={Agent-generated helper function for a batched indexing pattern with a flattened 4D array index, where multi-dimensional tensor indices $(batch,ch,h,w)$ are mapped into a single linear offset.},label=code4,frame=tlrb]
/*@
requires 0 <= batch && batch < batch_size && 0 <= ch && ch < channels;
requires 0 <= h && h < height && 0 <= w && w < width;
ensures \result == batch * channels * height * width
                 + ch * height * width + h * width + w;
ensures \result >= 0 && \result < batch_size * channels * height * width; @*/
/*@ pure @*/ int array_4d_index(..) =  
batch * channels * height * width + ch * height * width + h * width + w;

\end{lstlisting}

The entire VerCors Annotation Guide consists of approximately $\sim 9.5$K tokens and is also provided as part of the prompt along with the KnowledgeBase. While our guide is tailored to the domain of tensor algebra operators, our automatic iterative prompt refinement strategy can be generally applied to generate guides for other domains if the agent is provided with an initial set of hand-written examples to learn from.
\section{Agentic Semantic Equivalence Framework: Proving Functional Correctness}
\label{sec:roqagent}

This section presents an agentic approach to establishing semantic equivalence between the LLM-generated CUDA code and the original specification from which it was derived. A program is considered \textit{semantically equivalent} to its original specification if it faithfully preserves the intended behavior of the specification. In this work, we assume that the original specification is a PyTorch program. In order to determine whether the LLM-generated code matches the original specification, the specification must first be represented in a form amenable to mathematical reasoning. For this purpose, we leverage the Rocq Theorem Prover and its Gallina representation. 
Our semantic equivalence workflow can be summarized as follows:
\begin{itemize}
    \item First, we extract computational graph from Pytorch using our static analyzer (\S \ref{subsec:pytorchtool}).
    \item Then, we programmatically translate the computation graph into Rocq specification leveraging our MLRocq Library (\S \ref{subsec:mlrocq}).
    \item Next, the semantic equivalence agent synthesizes Rocq specifications from the CUDA kernel with additional pre and post-conditions as required.
    \item The semantic equivalence agent then constructs Rocq proofs to establish equivalence between the two Rocq specifications (\S \ref{subsec:rocqlowering}).
    \item Finally, the agent generates VerCors functional annotations from the Rocq specification (a step that is currently manually validated) and uses Vercors to make sure the CUDA kernel adheres to this functional specification under all conditions (\S \ref{subsec:rocqlowering}). 
\end{itemize}

\begin{figure*}[htbp]
\centering\includegraphics[width=1.0\textwidth]{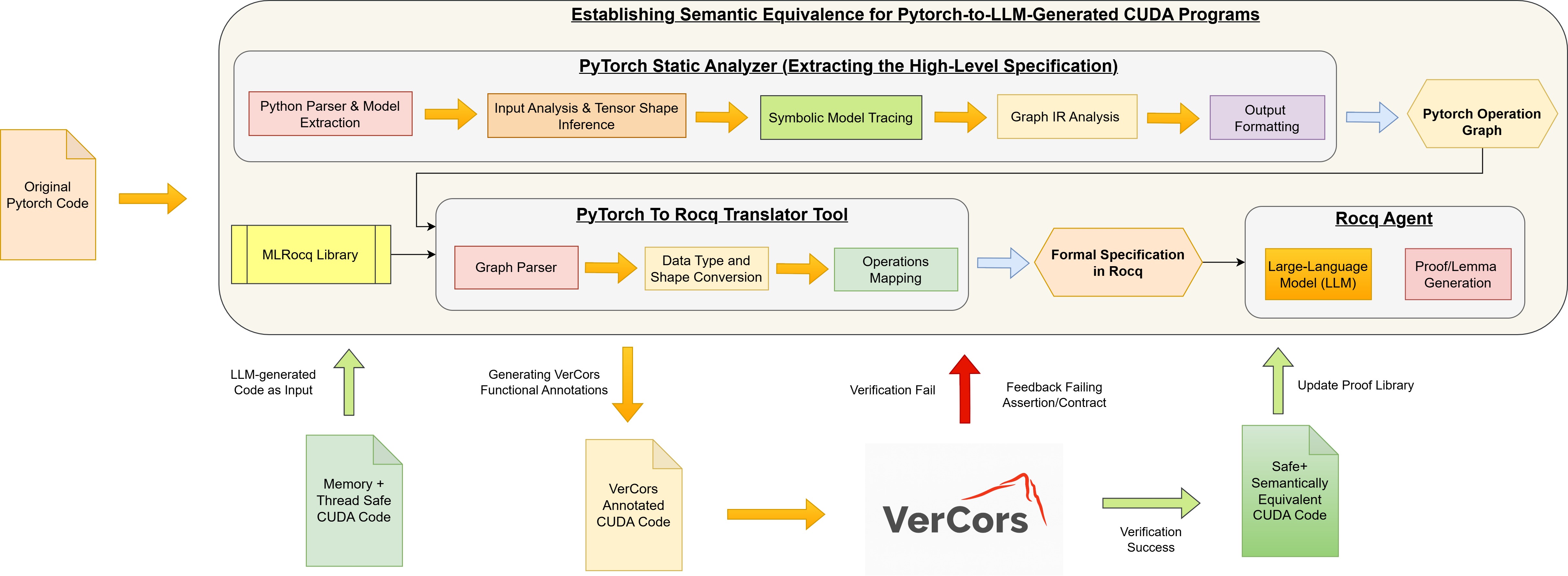}
\caption{\textbf{Agentic Semantic Equivalence Framework}. The \textit{front-end} captures the original specification in a graph-based IR form, and is translated to Rocq theorems. The agentic \textit{back-end} explores alternative mathematical representations of the original specification specification, and leverages them to generate low-level VerCors annotations. }
\label{fig:semantic1}
\end{figure*}


We describe the Semantic Equivalence framework in greater detail in the following sections.

\subsection{Front-End: Translating Pytorch Specification to Rocq}

The front-end of \textit{ProofWright}'s semantic equivalence framework (see Figure \ref{fig:semantic1}) consists of (1) the \textit{PyTorch Static Analyzer}, which captures the program's high-level specification (written in PyTorch) in a graph-based intermediate representation (IR), (2) the \textit{MLRocq library} which provides a corpus of formal definitions for popular neural and tensor computations in Rocq, and (3) the \textit{PyTorch to Rocq Translation} tool, which uses the MLRocq library to translate the extracted graph into equivalent Rocq representations. Note that the front end of the semantic equivalence framework does not employ any LLM components, ensuring that the formal specification process remains trustworthy.
\subsubsection{PyTorch Static Analyzer}
\label{subsec:pytorchtool}
\begin{figure*}[htbp]
\centering\includegraphics[width=0.9\textwidth]{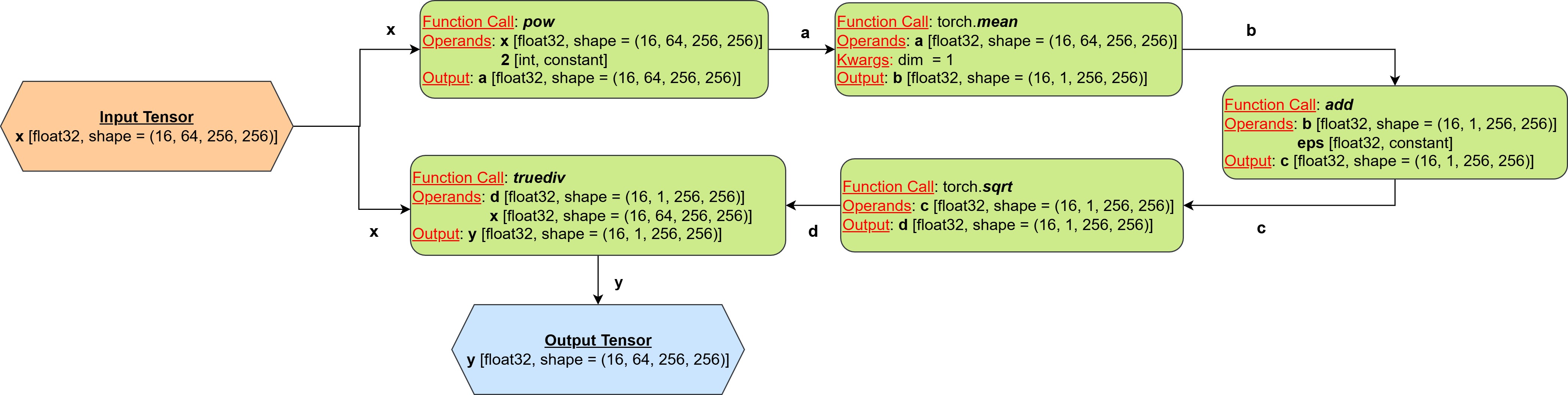}
\caption{\textbf{Operations Graph IR generated by Pytorch Static Analyzer for RMS}. The input tensor $x \in \mathbb{R}^{16 \times 64 \times 256 \times 256}$ is squared, reduced across the channel dimension, stabilized with an epsilon constant, square-rooted, and finally broadcast-divided with the original tensor to produce the normalized output.}
\label{fig:semantic2}
\end{figure*}

The \textit{PyTorch Static Analyzer} abstracts the high-level semantics of the pytorch input program by transforming it into a graph-based IR. Each computation in the PyTorch program is represented as an \textit{operation node}, while the edges capture data dependencies between operations, forming a dataflow graph that encodes the program’s computational structure. Each node in the graph is further enriched with tensor metadata, including shape information and auxiliary parameters such as dimensions and scaling factors. Figure \ref{fig:semantic2} shows an example of an operations graph for an RMS norm program from the KernelBench Benchmark. Internally, the tool leverages \textit{torch.fx} to symbolically trace the model, and record operations to construct an FX graph, which is then iteratively refined with tensor data (Kwargs, shape, etc) if possible. 

\subsubsection{MLRocq Library}
\label{subsec:mlrocq}

    
\begin{wrapfigure}{r}{0.45\textwidth}
\vspace{-10pt}
\begin{lstlisting}[language=Coq,caption={Definition of a recursive Coq type $TensorND$ that generalizes tensors to arbitrary dimensions.},label=code6,frame=tlrb]
Fixpoint TensorND (n:nat):Type:=
  match n with
| 0%nat => Z
| S n' => list (TensorND n')
\end{lstlisting}
\vspace{-20pt}
\end{wrapfigure}

\textit{MLRocq} is a formal verification library implemented in Rocq that provides mathematically rigorous definitions and proofs for deep learning operations. It defines a base data type $TensorND$, which represents an $n$-dimensional tensor, recursively constructed as a list of $(n-1)$-dimensional tensors 
The \textit{MLRocq} library currently implements nearly one hundred operations from the PyTorch library that are widely used in modern deep learning models. These include a broad spectrum of functional categories such as \textit{activation functions}, \textit{normalization} layers, \textit{pooling} and \textit{reduction} operations, and standard \textit{loss} functions. \footnote{A complete list of MLRocq operations is shown in Appendix A}.

\begin{lstlisting}[language=Coq,caption={Recursive definition of ReLU for n-d tensors which uses scalar version for base case and List.map for higher dimensions.},label=code7,frame=tlrb]
(* Element-wise ReLU: max(0, x) *)
Definition relu_sc (x:Z) : Z :=
  if x <? 0 then 0 else x.
(* ReLU for n-dim tensor *)
Fixpoint relu_t (n:nat) 
  (t:TensorND n) : (TensorND n) :=
  match n return TensorND n -> TensorND n with
  | 0%nat => fun x => relu_sc x
  | S n' => fun xs => 
    List.map (relu_TensorND n') xs
  end t.
\end{lstlisting}

Listing \ref{code7} shows an example where the ReLU activation is applied to an $n$-dimensional tensor. The function recursively applies the scalar ReLU to each element by mapping it across all sub-tensors using the \textit{map} operation, thereby extending element-wise activation to tensors of arbitrary dimensions. This represents a common implementation strategy in the \textit{MLRocq} library: each operation is first defined for the scalar case and then recursively applied to higher-dimensional tensors. Additionally, the library includes over 200 test cases designed to validate the correctness of these definitions with respect to their PyTorch equivalents. In its current version, most tensor and operation definitions are restricted to integer/real data types to simplify the construction of formal proofs.


\subsubsection{Pytorch-to-Roq Translator}
\label{subsec:rocqtrans}

The third component of the \textit{Semantic Equivalence Framework} is the translation stage, which takes the PyTorch operations graph and systematically maps each node to its formally defined Rocq counterpart using the MLRocq library. During this process, the translator resolves the operator with its corresponding mathematical definition (e.g., matrix multiplication, activations, etc) and lifts all operands into well-typed Rocq entities using tensor metadata (shape, type). This produces a formal, type-safe specification in Rocq that preserves the program’s original computational semantics and serves as the foundation for downstream theorem generation and equivalence proofs. Note that once a proof is completed, it is stored in the MLRocq library, in order to avoid recomputing again.

\subsection{Backend: Proving Semantic Equivalence}
The back-end of the semantic equivalence framework consists of the \textit{Semantic Equivalence Agent}, which generates the corresponding Rocq specification for the target CUDA program, builds a Rocq theorem, and automatically constructs proofs to validate its equivalence to the PyTorch specification. While this step proves that the LLM-generated Rocq representation of the CUDA program is equivalent to the original specification, we still do not know if the Rocq representation actually matches the CUDA program we are trying to verify. To this end, the agent lowers the verified specification into VerCors functional annotations, embeds them into the LLM-generated CUDA code, and invokes the VerCors tool to verify the correctness of the annotated program. 
\label{subsec:rocqlowering}
The final component of the \textit{Semantic Equivalence Framework} is the \textit{Semantic Equivalence Agent}, which operates on the formally translated specifications to establish semantic equivalence with the LLM-generated CUDA codes. It synthesizes Rocq theorems and auxiliary properties representing equivalent formulations of the CUDA program’s computational semantics and automatically constructs Rocq proofs to validate them against the original PyTorch specification.

As an example, starting from the original ReLU specification in Listing \ref{code7}, the \textit{Rocq Agent} synthesizes the corresponding Rocq representation of the CUDA program (Listing \ref{code9}), defining both the scalar base case and its recursive extension to $n$-dimensional tensors. In order to prove that these specifications are indeed equivalent, we observe that the agent automatically constructs the proof for each of these versions by performing a two-level inductive reasoning in a single shot (Listing \ref{code91}) and validates it using the Rocq tool.  

\begin{lstlisting}[language=Coq,caption={c},label=code9,frame=tlrb, caption = {Rocq representation of ReLU operation automatically synthesized by Semantic Equivalence Agent for scalar and n-dimensional tensor},float=htpb]
(* Scalar ReLU using arithmetic trick: (x + |x|) / 2 *)
Definition relu_arith (x : Z) : Z := (x + Z.abs x) / 2.
(* Extending to n-dimensional tensors *)
Fixpoint reluAR_t (n : nat) (t : TensorND n) : TensorND n :=
  match n return TensorND n -> TensorND n with
  | 0%nat => fun x => relu_arith x
  | S n' => fun xs => List.map (reluAR_t n') xs
  end t.  
\end{lstlisting}

\begin{lstlisting}[language=Coq,caption={c},label=code91,frame=tlrb, caption = {Rocq Agent Proving Equivalence between two diffent implmentations fo Relu}]
Theorem relu_scalar_equivalence: forall i: Z, relu_sc i = relu_arith i.
Proof. intros i. unfold relu_sc. unfold relu_arith.
       destruct (i <? 0) eqn:Hlt.
       + (*case i < 0*) apply Z.ltb_lt in Hlt. 
        rewrite Z_abs_neg by lia.
        rewrite Z.add_opp_r. rewrite Z.sub_diag. 
        rewrite Z.div_0_l by lia. reflexivity.
       + (*case i >= 0*) apply Z.ltb_ge in Hlt. 
         rewrite Z_abs_pos by lia. rewrite Z.add_diag. 
         symmetry. rewrite Z.mul_comm. apply Z.div_mul. lia.
Qed. 
(*Proving that two relu implementations are equivalent for N-d tensors*)
Theorem relu_NDtensor_eq: forall (n : nat) (t: TensorND n), relu_t n t = reluAR_t n t.
Proof.  induction n as [| n' IH]; intros t.
        + (* Base case: 0-dimensional tensor (Z) *) 
          apply relu_scalar_equivalence.
        + (* Inductive case: list of TensorND n' *) simpl. 
          induction t as [| x xs IHx].
          - simpl. reflexivity. 
          - simpl. rewrite IH. rewrite IHx. reflexivity.
Qed. 
\end{lstlisting}

The Rocq Agent proves that two different implementations of ReLU (Listings \ref{code7} \& \ref{code9}) are equivalent by reasoning over input cases ($i < 0$ and $i \geq 0$). It then generalizes this result to N-dimensional tensors using induction, showing that the equivalence holds for higher-dimensional tensors.

Finally, the agent derives the VerCors functional annotations from the combined space of original and synthesized specifications and embeds them into the LLM-generated GPU code for formal verification. A successful verification by VerCors indicates that the CUDA program has the intended output behavior and provides high assurance that it satisfies the original PyTorch specification. In contrast, if verification fails, the agent analyzes the error to infer the conditions under which the LLM-generated program deviates from the intended semantics. 

We note that our current approach does not guarantee the correctness of the lowering process that translates Rocq definitions into \textit{VerCors} functional annotations for all possible types of GPU Kernels because this step is performed using an LLM to demonstrate the feasibility of our approach. While we manually verified that the lowered VerCors annotations match the original specifications in our evaluation, in future work, we seek to replace this step with a procedural compiler.

\begin{lstlisting}[language=C++,frame=tlrb, caption = {VerCors Functional Annotation generated by Semantic Equivalence Agent.},label=code10]
// Functional correctness: x + |x|/2 operation
/*@
ensures \gtid<size ==> {:output[\gtid]:} == input[\gtid]+fabsf(input[\gtid])/2.0f;
@*/
__global__ void relu_kernel(float* input, float* output, int size) {
    int idx = blockIdx.x * blockDim.x + threadIdx.x;
    if (idx < size) {
        output[idx] = input[idx] + fabsf(input[idx]) / 2.0f;}}
\end{lstlisting}

\section{Evaluation}
\label{sec:eval}
The evaluation of \textit{ProofWright} is two-fold. The first set of experiments focuses on functional correctness, while the second set of experiments explore semantic equivalence of LLM-generated CUDA codes. Specifically, we design experiments to answer the following set of research questions:

\begin{itemize}
    \item \textbf{RQ1} (\S \ref{RQ1}): To what extent can the \textit{ProofWright} successfully establish memory and thread-safety guarantees for LLM-generated GPU codes? Furthermore, can it generalize to unseen kernels and handle increasingly complex verification scenarios? 
    \item \textbf{RQ2} (\S \ref{RQ2}): How lightweight is the annotation generation process, and does it scale to kernels that employ complex primitives and larger problem sizes?
    \item \textbf{RQ3} (\S \ref{RQ3}): How important are the individual components (\textit{knowledge base} and \textit{annotation guide}) in the \textit{VerCors Agent} prompt? Can the \textit{VerCors Agent} succeed in establishing memory and thread safety without them?
    \item \textbf{RQ4} (\S \ref{RQ4}): What kinds of high-level specifications can the semantic-equivalence framework capture and represent formally? Under what conditions is the \textit{Rocq Agent} able to synthesize alternate specification and automatically construct proofs? 
    \item \textbf{RQ5} (\S \ref{RQ5}): Finally, to what extent can \textit{ProofWright} establish semantic equivalence for LLM-generated GPU kernels, in spite of currently relying on an untrusted lowering stage?
    
\end{itemize}

\textbf{Experimental Setup}: The experiments were conducted on a MacBook Pro equipped with a 14-core Apple M4 Pro processor, running macOS Redwood. The \textit{VerCors} tool was built from source\footnote{\url{https://github.com/utwente-fmt/vercors}} and minimally extended to support additional CUDA constructs required for our experiments.

\textbf{Baselines}: The baseline CUDA kernels from KernelBench L1 \cite{ouyang2025kernelbenchllmswriteefficient} were generated using the Claude-4-Sonnet and GPT-5-Codex models. 

\textbf{Preprocessing for Vercors Compatibility}: To ensure compatibility with VerCors front-end, we made minimal type-compatibility changes (\texttt{float*} instead of \texttt{const float*}) to make kernels VerCors-compliant, including replacing unsupported headers with minimal CUDA interfaces (e.g., \texttt{\#include <cuda.h>}), and introduced pure function declarations for math operations.


\textbf{Setting up the Initial Annotation Guide}: The process of creating the initial version of annotation guide was as follows:
\begin{itemize}
    \item We took 10 manually annotated CUDA programs from the KnowledgeBase, and prompted the agent to invoke the VerCors tool and "\textit{generalize the learnings across these examples, and not pattern match}”.
    \item After each successful verification attempt, the agent was prompted to “\textit{identify any new insights/patterns not already present in the guide}” and update it accordingly.
   \item As an additional safety measure, all updates to the annotation guide were manually reviewed in the current implementation
\end{itemize}


\subsection{Memory \& Thread Safety: Extent of Coverage (RQ1)}
\label{RQ1}

\begin{figure*}[ht]
\begin{tabular}{p{0.5\textwidth}p{0.5\textwidth}}
    \begin{minipage}{.49\textwidth}
    \centering\includegraphics[width=1.0\textwidth]{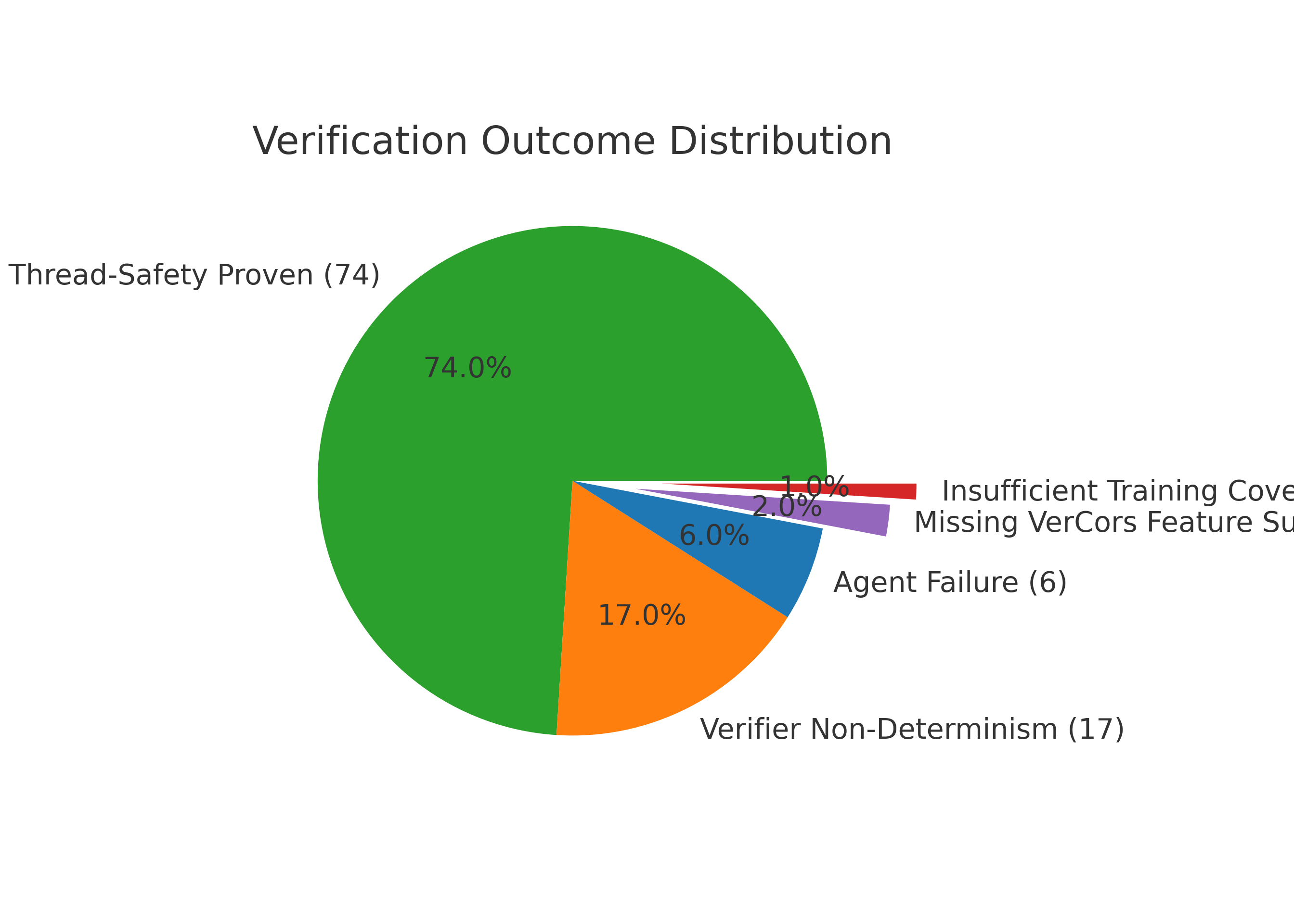}
    \caption{\small \textit{Verification outcome distribution for the Claude-4-Sonnet baseline on KernelBench L1}. The VerCors agent successfully established memory and thread-safety guarantees for the majority of kernels ($74\%$), while the remaining cases ($26\%$) exhibited verifier instability/agent limitations. Verifier instability mostly originated from non-affine indexing expressions ($15\%$)} 
    \label{fig:res-1}
    \end{minipage}
    &
    \begin{minipage}{.49\textwidth}
    \vspace{10pt}
    \centering\includegraphics[width=1.0\textwidth]{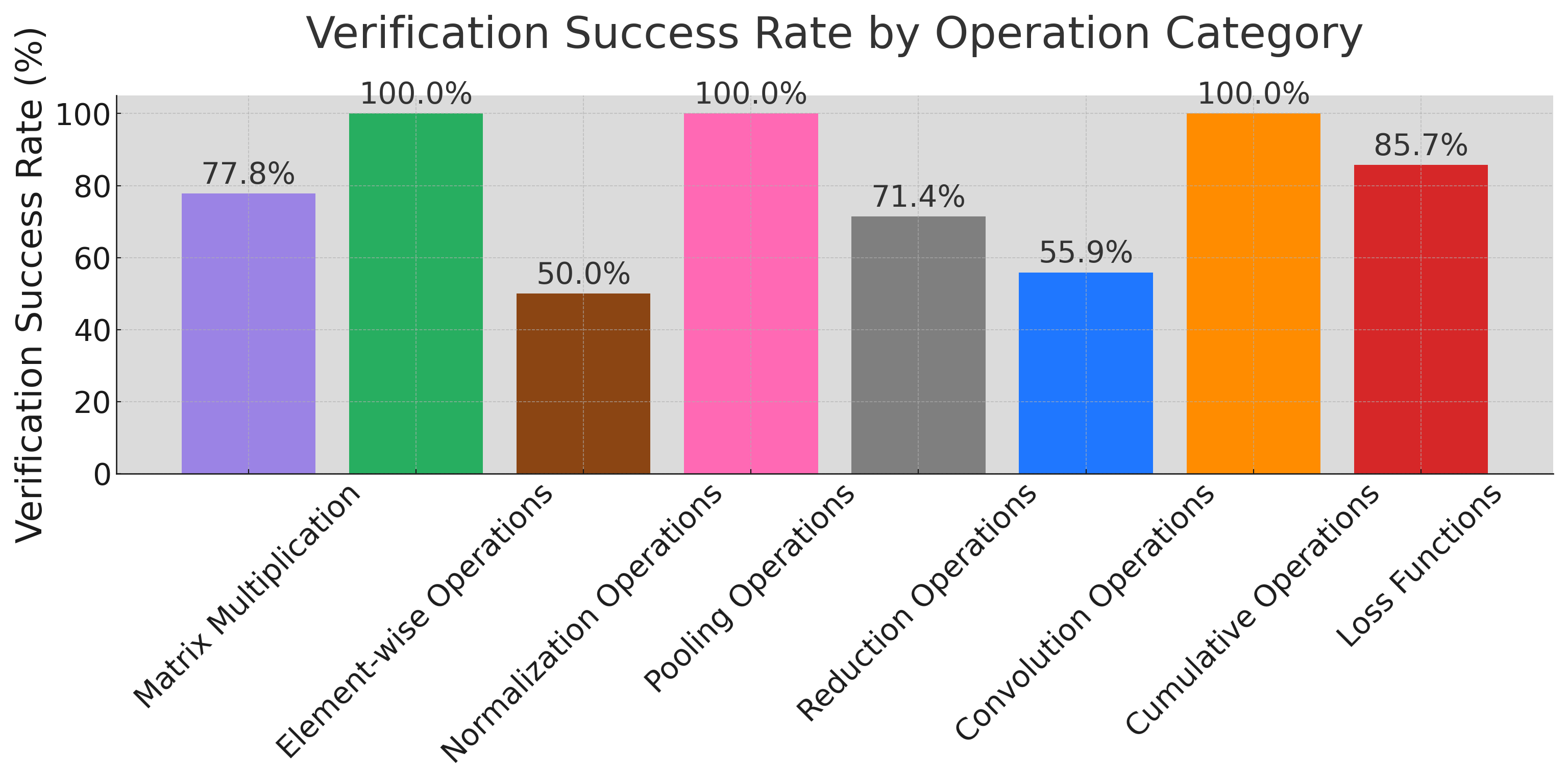}
    \caption{\small \textit{Verification coverage by operation category for the Claude-4-Sonnet baseline}. The VerCors agent achieved full verification for \textit{element-wise}, \textit{pooling}, and \textit{cumulative operations}, while matrix multiplication-based kernels exhibited partial coverage due to complex loop bounds and non-affine memory access patterns. Normalization and reduction operations showed moderate coverage.}
    \label{fig:res0}
    \end{minipage}
\end{tabular}
\end{figure*}

The \textit{VerCors Agent}, equipped with the \textit{knowledge base} and \textit{annotation guide} was tasked with generating annotations for the \textit{Claude-4-Sonnet} baseline across 100 L1 problems from the KernelBench benchmark. As shown in Fig.~\ref{fig:res-1}, the agent successfully established memory and thread safety for 74 kernels, spanning variants of \textit{matrix multiplication}, and other common deep-learning primitives. Fig.~\ref{fig:res0} further breaks down the success rate across different kernel categories. Overall, the outcomes of the \textit{VerCors Agent} can be grouped into three broad classes:

$\blacksquare$ \textit{Successful Verification} (74\%): In successfully verified cases, we observe that the agent is able to infer correct permissions for previously unseen kernels by generalizing permission patterns. It first classifies the kernel’s algorithmic family, analyzes its memory-access structure, and then selects an appropriate base annotation template, which is customized to the given kernel’s indexing and access patterns. For example, in the MaxPooling2D kernel (L1-42), the agent correctly recognized that each thread writes to a unique output index but performs overlapping reads, and generated the corresponding safety permission scheme. We also notice when an initial verification attempt fails, the agent decomposes complex, high-dimensional index expressions ($4D$ access) with helper functions ($2D$ accesses), which improves VerCors's ability to reason about them.

$\blacksquare$ \textit{Verifier Instability} (17\%): SMT solver instability is a known problematic phenomenon where seemingly insignificant changes to the program or specifications can cause program verification failure~\cite{zhou2023mariposa}. In cases where this instability was common (e.g., when specifications are quantified over multiple variables and include non-linear arithmetic), we found that the agent failed to verify programs but produced seemingly reasonable contracts and proof hints.
For instance, consider the following representative snippet derived from a real example that failed to verify:
\newpage
\begin{lstlisting}[language=C++,caption={A snippet from a failing verification attempt of correct code.},label=instable1,frame=tlrb,numbers=none]
//row = blockIdx.x * blockDim.x + threadIdx.x
//col = blockIdx.y * blockDim.y + threadIdx.y
for (int k = 0; k <= N - 4; k += 4) {
sum += A[k+row*N]     * B[col+k*N]       //Succeeds
    +  A[(k+1)+row*N] * B[col+(k+1)*N]   //Succeeds
    +  A[(k+2)+row*N] * B[col+(k+2)*N]   //Fails
    +  A[(k+3)+row*N] * B[col+(k+3)*N];  //Fails to prove 0 <= (k+3)+row*N < N*N
}
\end{lstlisting}
Note that removing the final two terms in the sum allows verification to succeed, yet it should succeed even if they are present, as all memory accesses are in-bounds. Through iterative debugging, we confirmed that this was indeed caused by solver instability.
Then we succeeded in manually verifying the kernel containing the above code
by adding the following \texttt{frame} statement that limited the SMT solver context:    
\begin{lstlisting}[language=C++,caption={Successful verification of Listing~\ref{instable1} after manually minimizing the solver context.},label=instable2,frame=tlrb, numbers=none]
for (int k = 0; k <= N - 4; k += 4) {
/*@ frame                                                                                                                
    requires 0 <= k && k <= N - 4;
    requires 0 <= row && row < N && 0 <= col && col < N;                                                                                          
    requires 0 <= threadIdx.x && threadIdx.x < blockDim.x;                                                                 
    requires 0 <= threadIdx.y && threadIdx.y < blockDim.y;                                                                 
    {  */ sum += A[k+row*N] * B[col+k*N] + ...    /*@ } */ 
} //Passes verificaiton
\end{lstlisting}

The underlying cause in this case is due to universal quantification over the \texttt{blockIdx} and \texttt{threadIdx} variables, which are unique to each thread; quantification over non-linear arithmetic is the primary cause of instability in our results.
SMT solver instability is a significant problem in the reliability of LLM agents for verification as it introduces more noise into the data available to the LLM and requires more debugging-style interaction with the verifier to differentiate true failures from unstable results.

$\blacksquare$ \textit{Agent Failure} ($9\%$): In a small subset of kernels, the agent fails to infer correct permission patterns even after multiple attempts. These cases typically arise from behaviors not generalizable from the agent’s current experiences, such as indirect addressing (e.g., $logits[logit\_idx]$ in L1-95.CrossEntropy), asymmetric or cooperative shared-memory loading, or limitations in VerCors support (e.g., handling inline structs). We expect these limitations to diminish as the agent is exposed to richer and more diverse training examples, and as the tool is further extended. 

\begin{tcolorbox}[enhanced,frame style image=blueshade.png,
  opacityback=0.75,opacitybacktitle=0.25,
  colback=blue!5!white,colframe=blue!75!black,
  title=Answer to RQ1: Coverage and Generalization]
  \textit{\textit{ProofWright} automatically establishes memory and thread-safety guarantees for 74\% of LLM-generated GPU kernels in KernelBench L1. The agent generalizes to previously unseen kernels by classifying algorithmic families and adapting annotation templates, while remaining failures are mainly due to unsupported non-affine patterns or SMT solver instability.}
\end{tcolorbox}

\subsection{Memory \& Thread Safety: Agentic Overheads (RQ2)}
\label{RQ2}

\begin{figure*}[htbp]
\centering\includegraphics[width=1.0\textwidth]{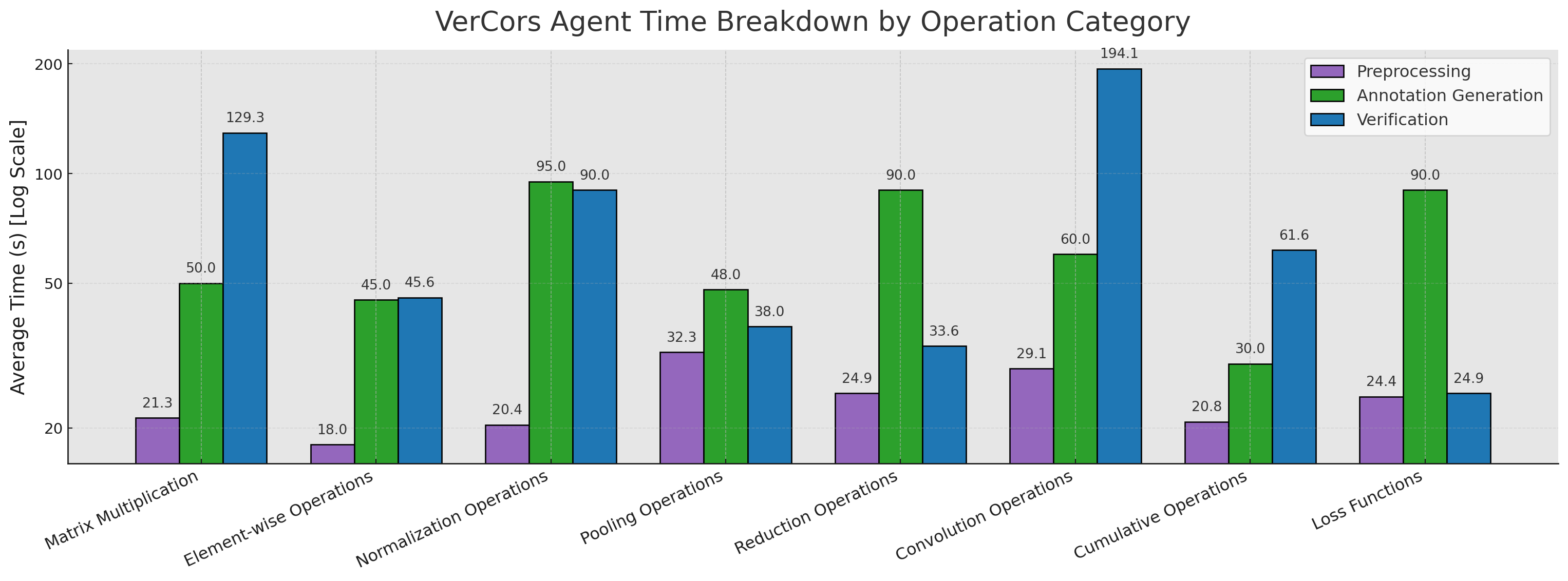}
\caption{\textit{Where does the VerCors Agent spends its time?} Average Agent time breakdown for the Claude-4-Sonnet baseline on KernelBench-L1. Most of the time was spent in verification (up to 194 s for convolution operations) and annotation generation ($\sim$90 s for reduction and normalization kernels), while preprocessing remained relatively lightweight across all categories.} 
\label{fig:res2}
\end{figure*}

As described in \S\ref{sec:vercorsagent}, the \textit{VerCors Agent} workflow consists of three major stages: a preprocessing phase, the annotation-generation phase, and the final verification stage. Like other agentic systems, the VerCors Agent operates in a feedback-driven loop, which implies that each failed verification introduces additional computational overheads as the agent iterates toward a satisfiable contract. Fig. \ref{fig:res2} summarizes the average time spent in each stage of this workflow.

The breakdown in Fig.~\ref{fig:res2} shows that verification dominates the overall runtime across \textit{matrix multiplications}, \textit{convolution} and \textit{cumulative} operations, since the verifier must statically reason about complex permission invariants and various thread interactions across the grid (especially pronounced in \textit{convolution operations}). However, an interesting trend is that certain categories, such as \textit{pooling}, \textit{reduction}, \textit{loss}, and \textit{normalization}, take longer to annotate than to verify. In these cases, the agent spends more time selecting and constructing hybrid permission templates for verification; particularly because these kernels were among the least represented in its knowledge base and annotation guide. Also, \textit{loss function} implementation often have more than one CUDA kernels, which attribute to high generation time. The appendix (\S C) further breaks down additional sources of overhead, including those arising from high-dimensional tensors, shared-memory usage, and other kernel-specific primitives.

\begin{tcolorbox}[enhanced,frame style image=blueshade.png,
  opacityback=0.75,opacitybacktitle=0.25,
  colback=blue!5!white,colframe=blue!75!black,
  title=Answer to RQ2: Agentic Overheads]
  \textit{ProofWright's annotation generation is lightweight ($\sim90$ s) with minimal preprocessing overhead, while overall runtime is dominated by verification (up to $194$ s), indicating the workflow scales reasonably for complex matmul kernels.}
\end{tcolorbox}

\subsection{Memory \& Thread Safety: Ablation Study (RQ3)}
\label{RQ3}

\begin{figure*}[htbp]
\centering\includegraphics[width=0.8\textwidth]{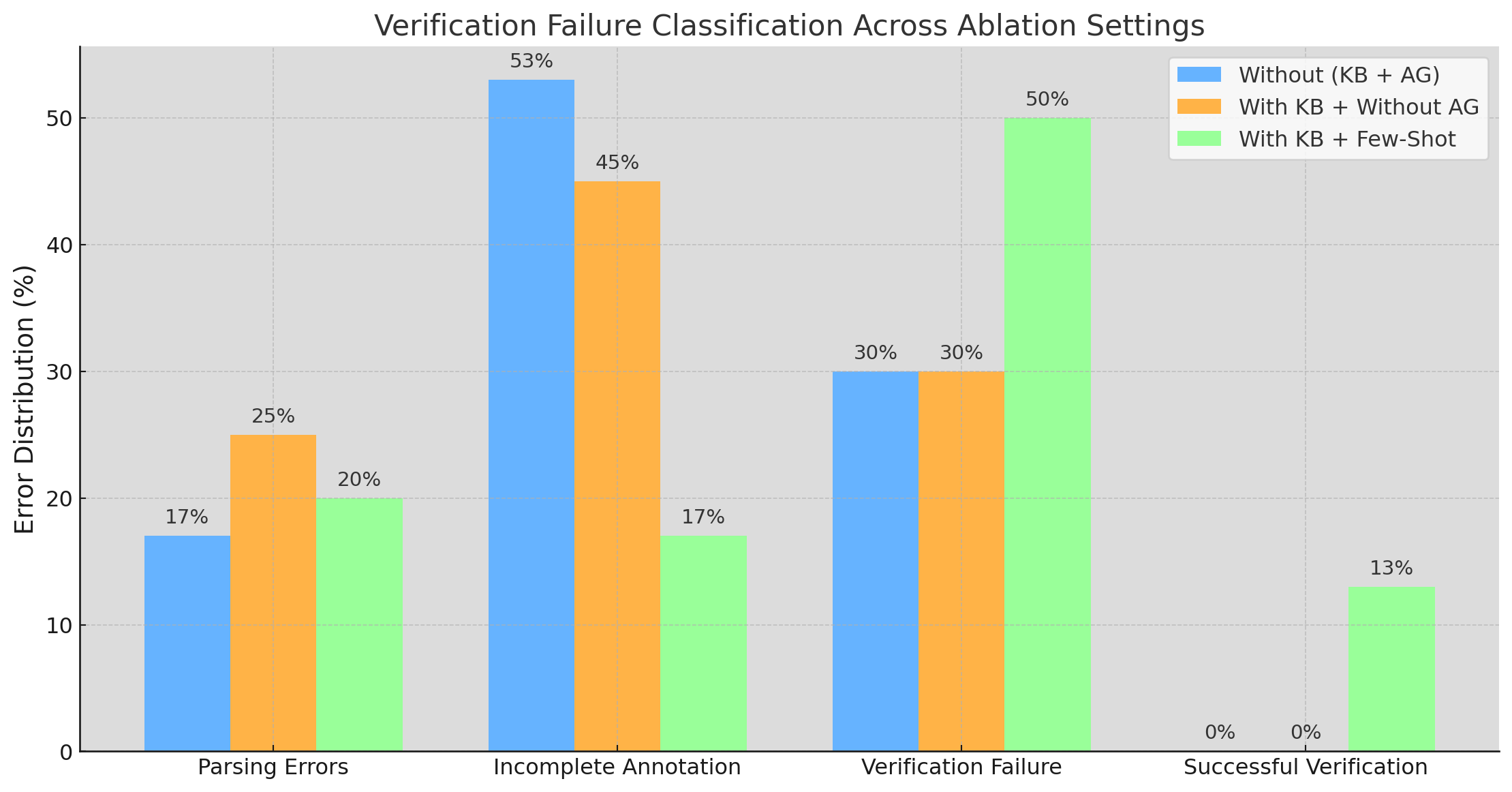}
    \caption{\small \textit{Are all the components essential in the VerCors Agent?} The agent was unable to verify even a single example in scenarios (1) \& (2), and could only verify 13 kernels in scenario (3)} 
    \label{fig:res3}
\end{figure*}

To further assess the contribution of the \textit{knowledge base} and the \textit{annotation guide} to the verification process, we conducted three experiments: (1) the agent was provided with neither the \textit{knowledge base} nor the \textit{annotation guide}, (2) the agent was given access only to the \textit{knowledge base} but not the \textit{annotation guide}, (3) The agent was given the knowledge base and 10 few-shot examples where were used to construct the initial annotation guide. Our experiments showed that \textit{VerCors} agent was \textit{unable to successfully verify even a single kernel in the first two scenarios}. To shed light into this extreme finding, we analyzed the underlying causes of failure in both execution modes (Fig.~\ref{fig:res3}).

We observed that in scenario (1), where neither the \textit{knowledge base} nor the \textit{annotation guide} is provided, the agent initially produces syntactically incorrect annotations, which it sometimes corrects after several attempts. However, even after syntactic repair, the resulting annotations fail to capture any meaningful thread-to-element reasoning and largely consist only of trivial conditions such as non-null pointer checks. In contrast, once the \textit{knowledge base} is introduced in scenario (2), the agent is able to generate reasonable annotations for some simpler kernels, but it failed to infer the enough weakest preconditions for verification (for instance $blockDim.x > 0$ instead of $blockDim.x == 32$). Listing 2 (Appendix \S C) presents a detailed example illustrating how annotations evolve as each component of the \textit{VerCors Agent} is incorporated.

In scenario (3), although the agent was able to verify a few examples, we found that it primarily relied on pattern matching rather than genuinely generalizing permission conditions. A further limitation was that, without the annotation guide acting as a form of long-term memory, the agent failed to accumulate or reuse insights from previously verified kernels, leading to repeated rediscovery of similar patterns rather than systematic learning.

\begin{tcolorbox}[enhanced,frame style image=blueshade.png,
  opacityback=0.75,opacitybacktitle=0.25,
  colback=blue!5!white,colframe=blue!75!black,
  title=Answer to RQ3: Importance of each component in ProofWright]
  \textit{Our ablation study shows that naive LLM-based integration via prompt engineering alone fails spectacularly: without the \textit{knowledge base} and \textit{annotation guide}, the agent verifies no kernels, and even with limited few-shot examples it verifies only 13\%, demonstrating that usefulness of  reusable verification knowledge through the persistent annotation guide.}
\end{tcolorbox}

\subsection{Semantic Equivalence: Capturing Functional Specification (RQ4)}
\label{RQ4}

The \textit{Semantic Equivalence} framework, which comprises of the \textit{static analysis tool}, \textit{mapping tool}, and \textit{Rocq Agent}, was tasked with extracting the functional specifications from the PyTorch programs in KernelBench, translating them into provably correct formal representations, and subsequently synthesizing the corresponding \textit{VerCors} functional annotations. The \textit{PyTorch Static Analyzer} successfully abstracted the high-level specifications for both the L1 and L2 problem sets, consisting of 200 Pytorch programs.

\begin{figure*}[ht]
\begin{tabular}{p{0.5\textwidth}p{0.5\textwidth}}
    \begin{minipage}{.49\textwidth}
    \centering\includegraphics[width=1.0\textwidth]{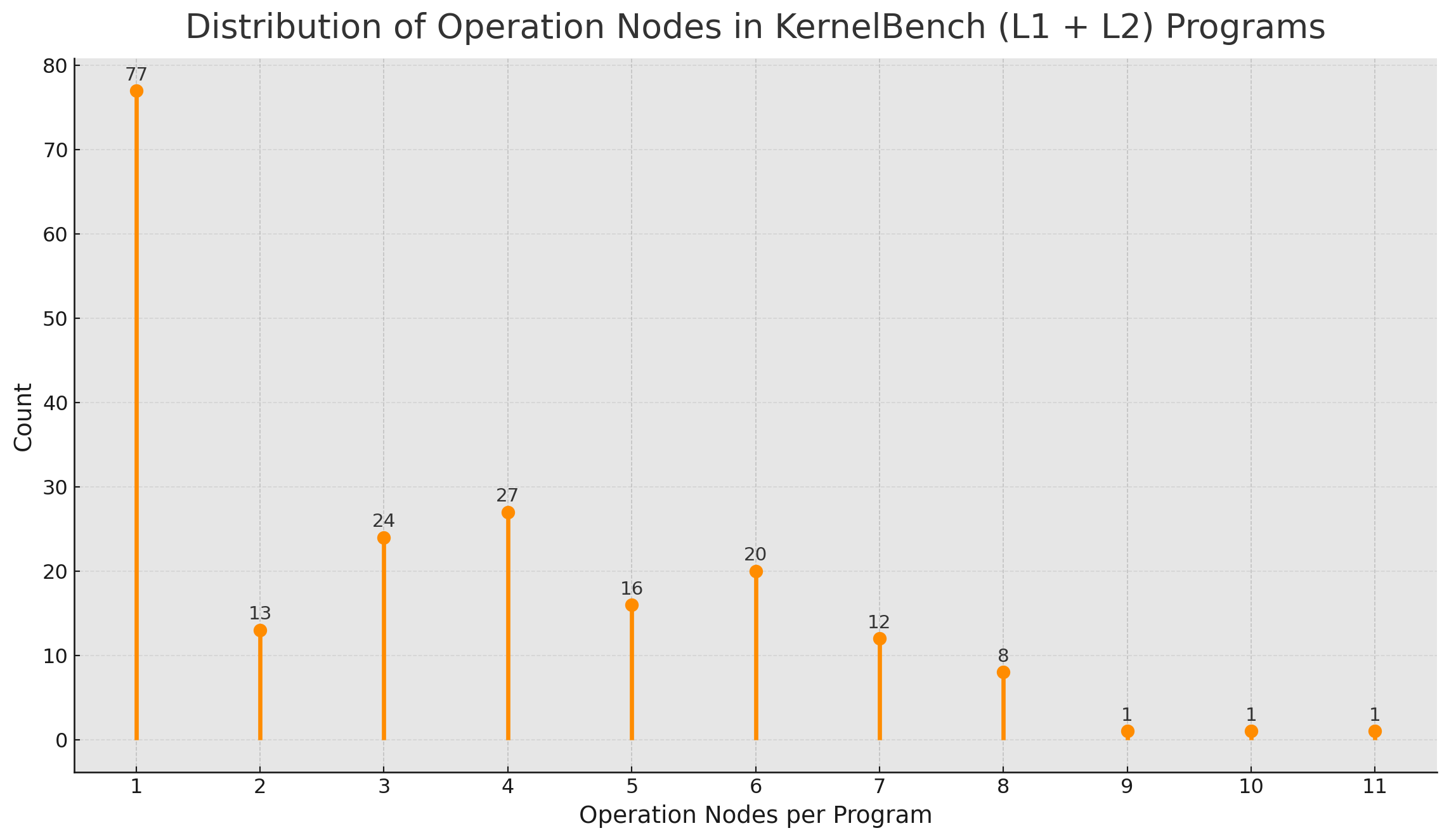}
    \caption{\small \textit{Distribution of Operation Nodes generated by Pytorch Static Analyzer}} 
    \label{fig:res4}
    \end{minipage}
    &
\begin{minipage}{.49\textwidth}
\centering
\resizebox{\columnwidth}{!}{%
\begin{tabular}{|c|l|c|}\hline
\textbf{Rank} & \textbf{PyTorch Operation} & \textbf{MLRocq Coverage} \\ \hline
1  & \texttt{torch.nn.Conv2d}          & $\checkmark$ \\ \hline
2  & \texttt{torch.sigmoid}            & $\checkmark$ \\ \hline
3  & \texttt{torch.nn.ConvTranspose3d} & $\times$      \\ \hline
4  & \texttt{torch.\_C.\_nn.gelu}      & $\checkmark$ \\ \hline
5  & \texttt{torch.mean}               & $\checkmark$ \\ \hline
6  & \texttt{torch.matmul}             & $\checkmark$ \\ \hline
7  & \texttt{torch.clamp}              & $\checkmark$ \\ \hline
8  & \texttt{gemm (torch.nn.Linear)}   & $\checkmark$ \\ \hline
9  & \texttt{torch.tanh}               & $\checkmark$ \\ \hline
10 & \texttt{torch.nn.GroupNorm}       & $\checkmark$ \\ \hline
\end{tabular}}
\caption{Top-10 Most Frequent PyTorch operations obtained from the graph IR, which are mapped into Rocq implementations.}
\label{tab:op}
\end{minipage}
\end{tabular}
\end{figure*}

Fig.~\ref{fig:res4} illustrates the distribution of nodes in the operations graphs generated by the \textit{PyTorch Static Analyzer} for KernelBench L1 and L2 programs. A larger number of nodes reflects a more complex functional specification, involving deeper dataflow structure and more computation steps that must ultimately be validated in the generated CUDA kernel. Currently, the \textit{MLRocq} library implements 99 of the PyTorch operations appearing in KernelBench Level~1, achieving approximately 92\% coverage. Table~\ref{tab:op} summarizes this by listing the most frequently occurring operations in KernelBench L1 and L2 and indicating whether they are currently supported in \textit{MLRocq}. The takeaway here is that the front end of the \textit{semantic equivalence framework} can extract precise formal specifications from PyTorch programs, and thus, can serve as a standalone specification abstraction layer.

\begin{tcolorbox}[enhanced,frame style image=blueshade.png,
  opacityback=0.75,opacitybacktitle=0.25,
  colback=blue!5!white,colframe=blue!75!black,
  title=Answer to RQ4: Capturing Functional Specification, 
  after skip=0pt]
  \textit{ProofWright captures functional specifications as operation-level computation graphs extracted from PyTorch programs and maps them into Rocq representations, covering $\sim$ 92\% of operations (99 ops) in KernelBench L1.}
\end{tcolorbox}

\subsection{Semantic Equivalence: Extent of ``Trusted" Coverage (RQ5)}
\label{RQ5}

\begin{wrapfigure}[18]{r}{0.45\textwidth}
\centering\includegraphics[width=0.5\textwidth]{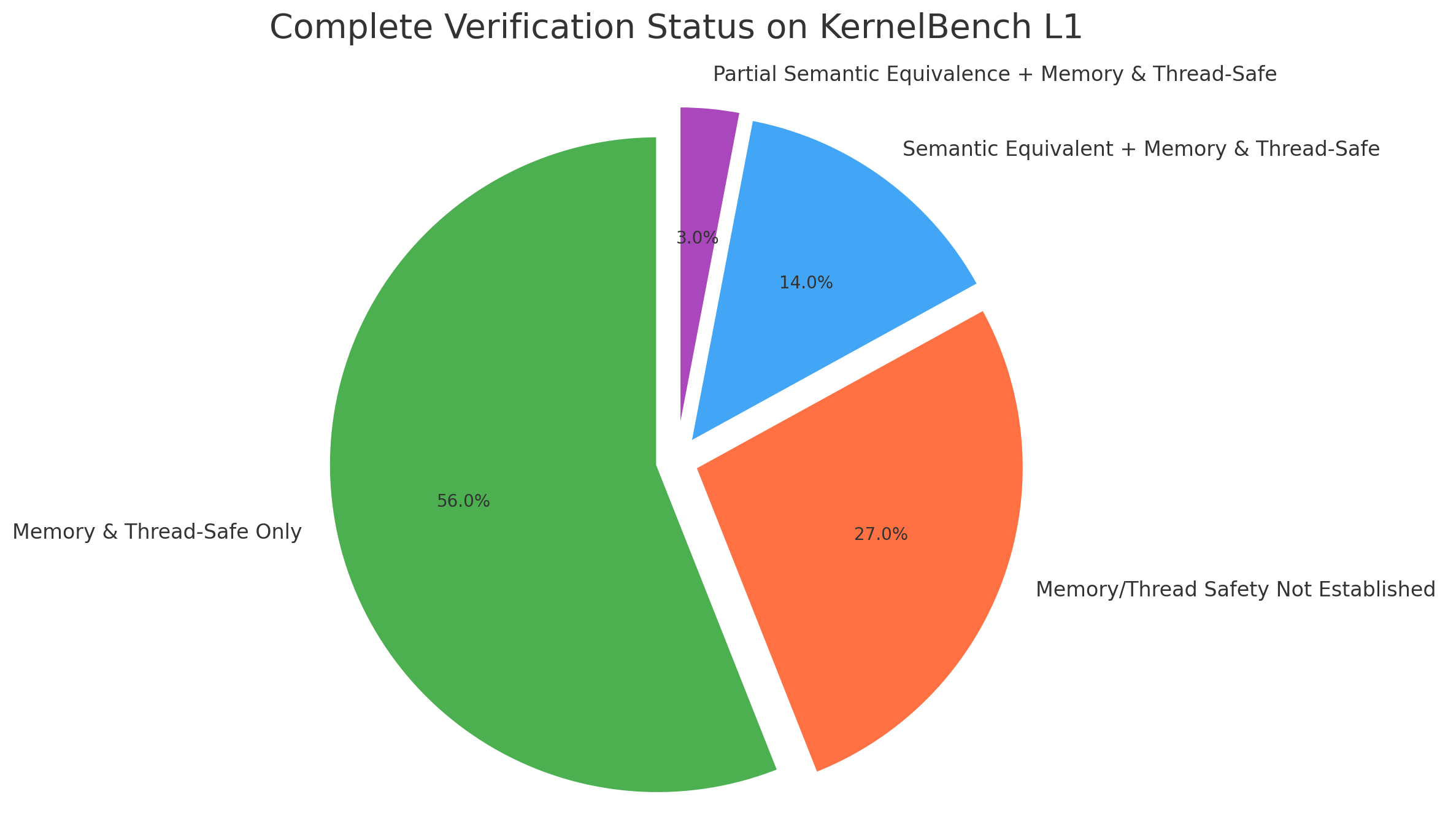}
    \caption{\small \textit{Distribution of Verification Guarantees for KernelBench L1 Programs.} Out of the 74\% memory and thread safe kernels, the semantic equivalence framework can establish functional correctness for 14\%  programs, while ensuring partial safety (one of out of two kernels verified) for an additional 3\% of the programs.} 
    \label{fig:res6}
\end{wrapfigure}

The high-level specification, together with the LLM-generated CUDA kernel, is passed to the \textit{Rocq Agent}, which synthesizes mathematical specifications from the CUDA implementation and automatically constructs proofs to validate they match the original PyTorch specification. These proven properties are then lowered into VerCors post-conditions to establish equivalence between the kernel's and the specification's outputs. Fig.~\ref{fig:res6} shows the final distribution of verified KernelBench L1 kernels. Note that VerCors requires memory and thread safety to be established before any functional correctness proof can succeed. The semantic equivalence framework was able to prove full correctness—including memory safety, thread safety, and semantic equivalence—for 14 CUDA kernels whose computation follows a simple one-to-one mapping pattern in which each GPU thread computes exactly one output element. These lowered annotations were manually validated to match the Rocq specification. This set of kernels primarily corresponds to activation-style operations within KernelBench, such as \textit{relu}, \textit{matrix-scalar multiplication}, etc. Listing \ref{sem1} shows an example (\textit{hardtanh}) where the \textit{semantic equivalence framework} transforms the VerCors functional annotations based on specifications in the MLRocq library. 

We also observed that the semantic equivalence framework was able to establish \textit{partial semantic equivalence} for three additional programs (\textit{MSELoss}, \textit{HuberLoss}, and \textit{HingeLoss}). Each of these CUDA programs contains two CUDA kernels: an element-wise kernel followed by a reduction kernel that aggregates intermediate results in shared memory. The agent successfully proved the functional correctness of the element-wise kernel, but was unable to establish the required postconditions for the subsequent reduction kernel. More broadly, the semantic equivalence framework struggles with kernels whose computations extend beyond simple one-to-one thread–element mappings—such as reductions, pooling, or any kernel where outputs depend on accumulated state, cross-thread interactions, or multi-step control flow. These patterns require expressing global invariants, loop-carried properties, and inter-thread reasoning, which exceed the capabilities of the current lowering logic used to translate mathematical specifications into VerCors postconditions.

\begin{lstlisting}[language=C++,caption={Rocq Agent proving semantic equivalence by adding mathemtical properties from the Rocq specifications of the HardTanh Function as functional post-conditions},label=sem1,frame=tlrb,numbers=none,float=htpb]
 /*@
// Functional correctness: HardTanh(x) = clamp(x, min_val, max_val)
  ensures \gtid < size ==> 
    {:output[\gtid]:} == (input[\gtid] < -1.0f ? -1.0f : (input[\gtid] > 1.0f ? 1.0f : input[\gtid]));
  // Mathematical properties from MLROCQ specification
  // Property 1: Bounded output range
  ensures \gtid < size ==> {:output[\gtid]:} >= -1.0f && {:output[\gtid]:} <= 1.0f;
  // Property 2: Identity within bounds (preserves values in [-1, 1])
  ensures \gtid < size && input[\gtid] >= -1.0f && input[\gtid] <= 1.0f ==> 
    {:output[\gtid]:} == input[\gtid];
  // Property 3: Lower bound saturation
  ensures \gtid < size && input[\gtid] < -1.0f ==> {:output[\gtid]:} == -1.0f;
  // Property 4: Upper bound saturation
  ensures \gtid < size && input[\gtid] > 1.0f ==> {:output[\gtid]:} == 1.0f;
@*/
__global__ void hardtanh_kernel(float* input, float* output, int size) {
    int idx = blockIdx.x * blockDim.x + threadIdx.x;
    if (idx < size) {
        float x = input[idx];
        float min_result = (x < HARDTANH_MAX) ? x : HARDTANH_MAX;
        output[idx] = (min_result > HARDTANH_MIN) ? min_result : HARDTANH_MIN;
    }}
\end{lstlisting}

\begin{tcolorbox}[enhanced,frame style image=blueshade.png,
  opacityback=0.75,opacitybacktitle=0.25,
  colback=blue!5!white,colframe=blue!75!black,
  title=Answer to RQ5: Extent of Trusted Coverage]
  \textit{\textit{ProofWright} establishes full semantic equivalence for 14\% of KernelBench L1 kernels, primarily element-wise kernels with one-to-one thread–output mappings, and partial equivalence for an additional 3\%. More complex kernels involving reductions or cross-thread dependencies remain challenging for the current lowering logic.}
\end{tcolorbox}

\section{Current Limitations}
\label{sec:limit}

While \textit{ProofWright} demonstrates promising results in verifying LLM-generated GPU kernels, it is currently limited by the subset of CUDA features supported by the underlying \textit{VerCors} verifier. At present, VerCors supports core constructs such as shared memory accesses and synchronization primitives (e.g., \texttt{\_syncthreads}), but does not yet model several modern CUDA features commonly used in high-performance kernels, including tensor core instructions ($mma, wgmma$). These limitations are not inherent to \textit{ProofWright} itself; the framework is agnostic to the verification backend and only requires a deductive verifier capable of reasoning about correctness through user-provided annotations. As support for additional CUDA primitives becomes available—either within VerCors or alternative verification frameworks—the \textit{annotation guide} can be readily extended, enabling \textit{ProofWright} to verify a broader class of GPU kernels.

\section{Related Work}
\label{sec:related}

\paragraph{GPU Program Verification}

There is a plethora of tools for GPU kernel verification that span the gamut from formal program verifiers to incomplete but low-overhead dynamic analysis~\cite{gpuformalsurvey}. Dynamic instrumentation mechanisms like Compute Sanitizer~\cite{nvidia2025computesanitizer} can check for unsafe behavior at runtime and symbolic execution tools like GKLEE~\cite{gklee2012} can strengthen the guarantees of typical dynamic analysis by making some parameters symbolic. However, these techniques are unsound because they do not consider all possible executions. In this work, we focus on program verifiers like GPUVerify~\cite{Betts2012GPUVerify} and VerCors that can provide sound guarantees about program behavior and which are guided by user annotations; these annotations provide an obvious interface for leveraging LLMs to automate verification. It is an open and interesting question as to how LLMs can be used in conjunction with other tools that do not expose a simple text interface for expert guidance.

\paragraph{LLM-based Program Verification}

Using LLMs to automate formal verification in deductive languages is an emerging field of interest. Several efforts~\cite{laurel2025,misu2024} have applied LLMs to generating or repairing Dafny programs and their specifications. Laurel ~\cite{laurel2025} uses an LLM to repair failing Dafny proofs within an existing codebase. They develop (1) heuristics to extract relevant examples from the codebase and (2) static analysis-based prompt strategies to improve the LLM success rate. While ProofWright could potentially benefit from (2), we cannot leverage the existence of a large codebase with proofs and programs relevant to the target. AutoVerus~\cite{autoverus} is an LLM-based system that generates proofs in Verus~\cite{verus2024}, a verification language for Rust which (like VerCors) suffers from low amounts of training data. AutoVerus takes as input a Rust program and Verus specification and guides an LLM to both generate a Verus proof and then iteratively repair failing proofs. AutoVerus bears many similarities to ProofWright barring two notable differences. First, Verus programs benefit from the guarantees of Rust's type system and do not need to prove memory safety or data race freedom of safe Rust fragments (putting less burden on the LLM to reason about concurrency or memory safety). Second, AutoVerus assumes that specifications are provided already in Verus; we demonstrate how to extract functional specifications from a high-level implementations.

\paragraph{Verifiable LLM Code Generation}
Some efforts seek a dual solution that tackles verification of LLM-generated code as part of the generation process. AlphaVerus~\cite{aggarwal2024alphaverus} translates both code and specifications from Dafny to Verus through an iterative search process over LLM-generated translations.
Other works~\cite{li2025dafnyverificationir,cloverllm} use LLMs to automatically generate specifications for programs from either natural language prompts or sets of test cases as part of LLM-based code generation. Although they increase confidence, these techniques still rely on a human or other entity to check that the specification matches the original intent. While ProofWright focuses purely on verification to separate the problems of code generation and verification, we imagine that the output of ProofWright could also be useful as feedback for LLM code generation tools.

\paragraph{LLM-based Interactive Proofs}
Interactive theorem provers like Rocq and Lean~\cite{mouralean2021} are used by a relatively small set of expert practitioners; therefore many groups~\cite{bayazit2025, wang2024legoprover, baldurfirst2023} have investigated the ability of LLMs to repair or generate proofs in these languages and have found promising initial results. Furthermore, proof abstraction and re-use has been shown to be an effective and automatic method for improving LLM proof generation capabilities overtime~\cite{baldurfirst2023}; this serves a similar purpose as the annotation guide in ProofWright. In this work, we demonstrate one potential use for LLM-aided interactive theorem proving: guiding automated program verifiers \emph{without} encoding the target program's semantics in the theorem prover's language.

\section{Conclusion}
\label{sec:conc}








In conclusion, ProofWright is an agentic CUDA verification framework that goes beyond traditional testing by providing robust guarantees for memory safety, thread safety, and semantic correctness. Conceptually, ProofWright demonstrates how prior knowledge (knowledge base), accumulated experience (annotation guide), and verifier feedback can be leveraged to enable automated verification of GPU kernels. Experiments on KernelBench L1 problems confirmed ProofWright's effectiveness, showing it verified thread and memory safety properties for up to 74\% of the kernels and demonstrating semantic equivalence for 14\% of the kernels with relatively low overheads and without the need for model training. ProofWright's methodology yielded crucial insights for building AI verification agents dealing with low-resource languages. Our findings show that additional components such as knowledge bases and annotation guides, are essential for generalization, while naive prompting or limited few-shot examples fail to generalize and often lead to pattern matching rather than correct reasoning. Moving forward, we plan to enhance our solution by extending CUDA language feature support in VerCors to support a larger set of CUDA programs, and developing an analytical Rocq-to-VerCors annotation translation tool to increase the reliability of our semantic equivalence framework.


\section{Data Availability Statement}
\textit{ProofWright} will be open-sourced, along with both its agentic flow and additional tools, and submitted for artifact evaluation.


\begin{acks}

\end{acks}

\bibliographystyle{ACM-Reference-Format}
\bibliography{sample-acmsmall}

@String{Computing = "Computing" }

@String{Computer = "{IEEE} Computer" }

@String{Springer = "Springer-Verlag" }

@inbook{2019pytorch,
author = {Paszke, Adam and Gross, Sam and Massa, Francisco and Lerer, Adam and Bradbury, James and Chanan, Gregory and Killeen, Trevor and Lin, Zeming and Gimelshein, Natalia and Antiga, Luca and Desmaison, Alban and K\"{o}pf, Andreas and Yang, Edward and DeVito, Zach and Raison, Martin and Tejani, Alykhan and Chilamkurthy, Sasank and Steiner, Benoit and Fang, Lu and Bai, Junjie and Chintala, Soumith},
title = {PyTorch: an imperative style, high-performance deep learning library},
year = {2019},
booktitle = {Proceedings of the 33rd International Conference on Neural Information Processing Systems},
articleno = {721},
numpages = {12},
doi= {10.5555/3454287.3455008}
}

@article{damani2024warpdrive,
  title={Warpdrive: An agentic workflow for ninja gpu transformations},
  author={Damani, Sana and Hari, Siva Kumar Sastry and Stephenson, Mark and Kozyrakis, Christos},
  year={2024}
}

@misc{lange2025aicuda,
    title   = {The AI CUDA Engineer: Agentic CUDA Kernel Discovery, Optimization and Composition},
    author  = {Lange, Robert Tjarko and Prasad, Aaditya and Sun, Qi and Faldor, Maxence and Tang, Yujin and Ha, David},
    year    = {2025},
    month   = {February},
    howpublished = {Sakana AI Blog Post},
    url     = {https://sakana.ai/ai-cuda-engineer/}
}

@inproceedings{Betts2012GPUVerify,
  author    = {Adam Betts and Nathan Chong and Alastair F. Donaldson and Shaz Qadeer and Paul Thomson},
  title     = {{GPUVerify: A Verifier for GPU Kernels}},
  booktitle = {Proceedings of the ACM International Conference on Object Oriented Programming Systems Languages and Applications, OOPSLA 2012, part of SPLASH 2012, Tucson, AZ, USA, October 21-25, 2012},
  year      = {2012},
  pages     = {133-152},
  publisher = {{ACM}},
  doi       = {10.1145/2384616.2384628},
  note      = {Addresses "two classes of bugs which make writing correct GPU kernels harder than writing correct sequential code: data races and barrier divergence." [2]}
}

@misc{baronio2025kevin32b,
  author       = {Carlo Baronio and Pietro Marsella and Ben Pan and Silas Alberti},
  title        = {Kevin-32B: Multi-Turn RL for Writing CUDA Kernels},
  howpublished = {\url{https://cognition.ai/blog/kevin-32b}},
  note         = {Cognition AI Blog},
  year         = {2025},
  month        = {May},
  day          = {6}
}

@inproceedings{cummins2025llm,
  title={LLM Compiler: Foundation Language Models for Compiler Optimization},
  author={Cummins, Chris and Seeker, Volker and Grubisic, Dejan and Roziere, Baptiste and Gehring, Jonas and Synnaeve, Gabriel and Leather, Hugh},
  booktitle={Proceedings of the 34th ACM SIGPLAN International Conference on Compiler Construction},
  pages={141--153},
  year={2025}
}

@inproceedings{taneja2025llm,
  title={Llm-vectorizer: Llm-based verified loop vectorizer},
  author={Taneja, Jubi and Laird, Avery and Yan, Cong and Musuvathi, Madan and Lahiri, Shuvendu K},
  booktitle={Proceedings of the 23rd ACM/IEEE International Symposium on Code Generation and Optimization},
  pages={137--149},
  year={2025}
}

@article{huang2024effilearner,
  title={Effilearner: Enhancing efficiency of generated code via self-optimization},
  author={Huang, Dong and Dai, Jianbo and Weng, Han and Wu, Puzhen and Qing, Yuhao and Cui, Heming and Guo, Zhijiang and Zhang, Jie},
  journal={Advances in Neural Information Processing Systems},
  volume={37},
  pages={84482--84522},
  year={2024}
}

@article{peng2024perfcodegen,
  title={PerfCodeGen: Improving Performance of LLM Generated Code with Execution Feedback},
  author={Peng, Yun and Gotmare, Akhilesh Deepak and Lyu, Michael and Xiong, Caiming and Savarese, Silvio and Sahoo, Doyen},
  journal={arXiv preprint arXiv:2412.03578},
  year={2024}
}

@inproceedings{Li2010ScalableSMT,
  author    = {Guodong Li and Sriraman Sankar and Ganesh Gopalakrishnan and Indradeep Ghosh and Sreeranga P. Rajan},
  title     = {Scalable SMT-Based Verification of GPU Kernel Functions},
  booktitle = {Proceedings of the 18th ACM SIGSOFT International Symposium on Foundations of Software Engineering (FSE '10)},
  year      = {2010},
  publisher = {ACM},
  address   = {Santa Fe, New Mexico, USA},
  pages     = {97--106},
  doi       = {10.1145/1882291.1882305},
  note      = {Page numbers and DOI are standard for FSE papers and typically sourced from ACM Digital Library; [2] confirms other details.}
}

@inproceedings{Li2012Parameterized,
  author    = {Guodong Li and Ganesh Gopalakrishnan},
  title     = {Parameterized Verification of GPU Kernel Programs},
  booktitle = {2012 IEEE 26th International Parallel and Distributed Processing Symposium Workshops \& PhD Forum (IPDPSW)},
  year      = {2012},
  month     = {May},
  pages     = {2149--2158},
  doi       = {10.1109/IPDPSW.2012.302},
  publisher = {IEEE},
  note      = {Details primarily from.[3] Page numbers sourced via IEEE Xplore using the DOI.}
}

@inproceedings{Li:2012:GKV:2145816.2145823,
  author    = {Li, Guodong and Li, Peng and Sawaya, Geof and Gopalakrishnan, Ganesh and Ghosh, Indradeep and Rajan, Sreeranga P.},
  title     = {{GKLEE: Concolic Verification and Test Generation for GPUs}},
  booktitle = {Proceedings of the 17th ACM SIGPLAN Symposium on Principles and Practice of Parallel Programming},
  series    = {PPoPP '12},
  year      = {2012},
  isbn      = {978-1-4503-1160-1},
  location  = {New Orleans, Louisiana, USA},
  pages     = {11--20},
  numpages  = {10},
  url       = {http://doi.acm.org/10.1145/2145816.2145823},
  doi       = {10.1145/2145816.2145823},
  acmid     = {2145823},
  publisher = {ACM},
  address   = {New York, NY, USA}
}

@inproceedings{Berkovich2013GPUbased,
  author    = {Shay Berkovich and Borzoo Bonakdarpour and Sebastian Fischmeister},
  title     = {{GPU-based Runtime Verification}},
  booktitle = {2013 IEEE International Parallel \& Distributed Processing Symposium (IPDPS)},
  year      = {2013},
  month     = {May},
  pages     = {747--758},
  doi       = {10.1109/IPDPS.2013.73},
  publisher = {IEEE},
  note      = {Details primarily from.[6, 7] Page numbers and DOI sourced via IEEE Xplore.}
}

@inproceedings{ZhengRLDS15,
  author    = {Manchun Zheng and
               Michael S. Rogers and
               Ziqing Luo and
               Matthew B. Dwyer and
               Stephen F. Siegel},
  title     = {{CIVL: Formal Verification of Parallel Programs}},
  booktitle = {30th IEEE/ACM International Conference on Automated Software Engineering, {ASE} 2015, Lincoln, NE, USA, November 9-13, 2015},
  pages     = {830--835},
  publisher = {{IEEE} Computer Society},
  year      = {2015},
  url       = {https://doi.org/10.1109/ASE.2015.99},
  doi       = {10.1109/ASE.2015.99},
  bibsource = {dblp computer science bibliography, https://dblp.org}
}

@misc{ouyang2025kernelbenchllmswriteefficient,
      title={KernelBench: Can LLMs Write Efficient GPU Kernels?}, 
      author={Anne Ouyang and Simon Guo and Simran Arora and Alex L. Zhang and William Hu and Christopher Ré and Azalia Mirhoseini},
      year={2025},
      eprint={2502.10517},
      archivePrefix={arXiv},
      primaryClass={cs.LG},
      url={https://arxiv.org/abs/2502.10517}, 
}

@inproceedings{blom2014vercors,
  author    = {Stefan Blom and Marieke Huisman},
  title     = {The VerCors Tool for Verification of Concurrent Programs},
  booktitle = {Proceedings of the 19th International Symposium on Formal Methods (FM 2014)},
  editor    = {Cliff Jones and Pekka Pihlajasaari and Jun Sun},
  series    = {Lecture Notes in Computer Science},
  volume    = {8442},
  pages     = {127--131},
  publisher = {Springer},
  address   = {Berlin, Germany},
  year      = {2014},
  doi       = {10.1007/978-3-319-06410-9\_9},
  url       = {https://doi.org/10.1007/978-3-319-06410-9\_9}
}

@inproceedings{leino2010dafny,
  title={Dafny: An automatic program verifier for functional correctness},
  author={Leino, K Rustan M},
  booktitle={International conference on logic for programming artificial intelligence and reasoning},
  pages={348--370},
  year={2010},
  organization={Springer},
  doi={10.1007/978-3-642-17511-4\_20}
}

@inproceedings{swamy2016fstar,
    author = {Swamy, Nikhil and Hri\c{t}cu, C\u{a}t\u{a}lin and Keller, Chantal and Rastogi, Aseem and Delignat-Lavaud, Antoine and Forest, Simon and Bhargavan, Karthikeyan and Fournet, C\'{e}dric and Strub, Pierre-Yves and Kohlweiss, Markulf and Zinzindohoue, Jean-Karim and Zanella-B\'{e}guelin, Santiago},
    title = {Dependent types and multi-monadic effects in F*},
    year = {2016},
    isbn = {9781450335492},
    publisher = {Association for Computing Machinery},
    address = {New York, NY, USA},
    url = {https://doi.org/10.1145/2837614.2837655},
    doi = {10.1145/2837614.2837655}
}

@manual{nvidia2025computesanitizer,
  title        = {Compute Sanitizer Documentation},
  author       = {{NVIDIA Corporation}},
  year         = {2025},
  month        = {April},
  note         = {Version 2025.2.0},
  url          = {https://docs.nvidia.com/compute-sanitizer/ComputeSanitizer/index.html}
}

@inproceedings{Ferrell2019CUDA,
  title        = {{CUDA au Coq: A Framework for Machine-validating GPU Assembly Programs}},
  author       = {Benjamin Ferrell and Jun Duan and Kevin W. Hamlen},
  booktitle    = {Design, Automation And Test in Europe (DATE 2019)},
  pages        = {474--479},
  year         = {2019},
  organization = {EDAA},
  doi          = {10.23919/DATE.2019.8715160}
}

@inproceedings{armborst2024vercors,
  title={The VerCors verifier: a progress report},
  author={Armborst, Lukas and Bos, Pieter and van den Haak, Lars B and Huisman, Marieke and Rubbens, Robert and {\c{S}}akar, {\"O}mer and Tasche, Philip},
  booktitle={International Conference on Computer Aided Verification},
  pages={3--18},
  year={2024},
  organization={Springer}
}

@article{haack2015permission,
  title={Permission-based separation logic for multithreaded Java programs},
  author={Haack, Christian and Huisman, Marieke and Hurlin, Cl{\'e}ment and Amighi, Afshin},
  journal={Logical Methods in Computer Science},
  volume={11},
  year={2015},
  publisher={Episciences. org}
}

@software{NVIDIA_Compute_Sanitizer,
  title        = {NVIDIA Compute Sanitizer Tools \&\, API},
  author       = {{NVIDIA Corporation}},
  year         = {2025},
  url          = {https://developer.nvidia.com/compute-sanitizer},
  note         = {Accessed: 2025-Oct-28}
}

@article{peng2018curd,
  title={CURD: a dynamic CUDA race detector},
  author={Peng, Yuanfeng and Grover, Vinod and Devietti, Joseph},
  journal={ACM SIGPLAN Notices},
  volume={53},
  number={4},
  pages={390--403},
  year={2018},
  publisher={ACM New York, NY, USA}
}

@article{tarek2023cucatch,
  title={Cucatch: A debugging tool for efficiently catching memory safety violations in cuda applications},
  author={Tarek Ibn Ziad, Mohamed and Damani, Sana and Jaleel, Aamer and Keckler, Stephen W and Stephenson, Mark},
  journal={Proceedings of the ACM on Programming Languages},
  volume={7},
  number={PLDI},
  pages={124--147},
  year={2023},
  publisher={ACM New York, NY, USA}
}

@inproceedings{kamath2021iguard,
  title={Iguard: In-gpu advanced race detection},
  author={Kamath, Aditya K and Basu, Arkaprava},
  booktitle={Proceedings of the ACM SIGOPS 28th Symposium on Operating Systems Principles},
  pages={49--65},
  year={2021}
}

@inproceedings{kamath2020scord,
  title={ScoRD: A scoped race detector for GPUs},
  author={Kamath, Aditya K and George, Alvin A and Basu, Arkaprava},
  booktitle={2020 ACM/IEEE 47th Annual International Symposium on Computer Architecture (ISCA)},
  pages={1036--1049},
  year={2020},
  organization={IEEE}
}

@article{jacobson2024hirace,
  title={HiRace: Accurate and Fast Source-Level Race Checking of GPU Programs},
  author={Jacobson, John and Burtscher, Martin and Gopalakrishnan, Ganesh},
  journal={arXiv preprint arXiv:2401.04701},
  year={2024}
}

@article{laurel2025,
author = {Mugnier, Eric and Gonzalez, Emmanuel Anaya and Polikarpova, Nadia and Jhala, Ranjit and Yuanyuan, Zhou},
title = {Laurel: Unblocking Automated Verification with Large Language Models},
year = {2025},
publisher = {Association for Computing Machinery},
number = {OOPSLA1},
url = {https://doi.org/10.1145/3720499},
doi = {10.1145/3720499},
journal = {Proc. ACM Program. Lang.}
}

@article{misu2024,
author = {Misu, Md Rakib Hossain and Lopes, Cristina V. and Ma, Iris and Noble, James},
title = {Towards AI-Assisted Synthesis of Verified Dafny Methods},
year = {2024},
publisher = {Association for Computing Machinery},
number = {FSE},
url = {https://doi.org/10.1145/3643763},
doi = {10.1145/3643763},
journal = {Proc. ACM Softw. Eng.}
}

@inproceedings{verus2024,
author = {Lattuada, Andrea and Hance, Travis and Bosamiya, Jay and Brun, Matthias and Cho, Chanhee and LeBlanc, Hayley and Srinivasan, Pranav and Achermann, Reto and Chajed, Tej and Hawblitzel, Chris and Howell, Jon and Lorch, Jacob R. and Padon, Oded and Parno, Bryan},
title = {Verus: A Practical Foundation for Systems Verification},
year = {2024},
isbn = {9798400712517},
publisher = {Association for Computing Machinery},
doi = {10.1145/3694715.3695952},
booktitle = {Proceedings of the ACM SIGOPS 30th Symposium on Operating Systems Principles}
}

@article{autoverus,
author = {Yang, Chenyuan and Li, Xuheng and Misu, Md Rakib Hossain and Yao, Jianan and Cui, Weidong and Gong, Yeyun and Hawblitzel, Chris and Lahiri, Shuvendu and Lorch, Jacob R. and Lu, Shuai and Yang, Fan and Zhou, Ziqiao and Lu, Shan},
title = {AutoVerus: Automated Proof Generation for Rust Code},
year = {2025},
publisher = {Association for Computing Machinery},
number = {OOPSLA2},
url = {https://doi.org/10.1145/3763174},
doi = {10.1145/3763174},
journal = {Proc. ACM Program. Lang.}
}

@misc{aggarwal2024alphaverus,
      title={AlphaVerus: Bootstrapping Formally Verified Code Generation through Self-Improving Translation and Treefinement}, 
      author={Pranjal Aggarwal and Bryan Parno and Sean Welleck},
      year={2024},
      eprint={2412.06176},
      archivePrefix={arXiv},
      primaryClass={cs.LG},
      url={https://arxiv.org/abs/2412.06176}, 
}

@misc{li2025dafnyverificationir,
      title={Dafny as Verification-Aware Intermediate Language for Code Generation}, 
      author={Yue Chen Li and Stefan Zetzsche and Siva Somayyajula},
      year={2025},
      eprint={2501.06283},
      archivePrefix={arXiv},
      primaryClass={cs.SE},
      url={https://arxiv.org/abs/2501.06283}, 
}

@inproceedings{cloverllm,
author = {Sun, Chuyue and Sheng, Ying and Padon, Oded and Barrett, Clark},
title = {Clover: Closed-Loop Verifiable Code Generation},
year = {2024},
isbn = {978-3-031-65111-3},
publisher = {Springer-Verlag},
url = {https://doi.org/10.1007/978-3-031-65112-0_7},
doi = {10.1007/978-3-031-65112-0_7},
booktitle = {AI Verification: First International Symposium, SAIV 2024, Montreal, QC, Canada, July 22–23, 2024, Proceedings}
}

@inproceedings{gpuformalsurvey,
  title={Formal methods for gpgpu programming: Is the demand met?},
  author={van den Haak, Lars B and Wijs, Anton and van den Brand, Mark and Huisman, Marieke},
  booktitle={International Conference on Integrated Formal Methods},
  pages={160--177},
  year={2020},
  organization={Springer}
}

@article{gklee2012,
author = {Li, Guodong and Li, Peng and Sawaya, Geof and Gopalakrishnan, Ganesh and Ghosh, Indradeep and Rajan, Sreeranga P.},
title = {GKLEE: concolic verification and test generation for GPUs},
year = {2012},
publisher = {Association for Computing Machinery},
address = {New York, NY, USA},
url = {https://doi.org/10.1145/2370036.2145844},
doi = {10.1145/2370036.2145844},
journal = {SIGPLAN Not.}
}

@INPROCEEDINGS{aspetakis2024,
  author={Aspetakis, Dimitris and Kosmidis, Leonidas and Trompouki, Matina Maria and Ruiz, Jose and Marosy, Gabor},
  booktitle={2024 Design, Automation \& Test in Europe Conference \& Exhibition (DATE)}, 
  title={Formal Methods for High Integrity GPU Software Development and Verification}, 
  year={2024},
  doi={10.23919/DATE58400.2024.10546867}
}

@inproceedings{zhou2023mariposa,
  title={Mariposa: Measuring SMT instability in automated program verification},
  author={Zhou, Yi and Bosamiya, Jay and Takashima, Yoshiki and Li, Jessica and Heule, Marijn and Parno, Bryan},
  booktitle={2023 Formal Methods in Computer-Aided Design (FMCAD)},
  year={2023},
  organization={IEEE},
  doi={10.34727/2023/isbn.978-3-85448-060-0_26}
}

@inproceedings{baldurfirst2023,
author = {First, Emily and Rabe, Markus N. and Ringer, Talia and Brun, Yuriy},
title = {Baldur: Whole-Proof Generation and Repair with Large Language Models},
year = {2023},
isbn = {9798400703270},
publisher = {Association for Computing Machinery},
url = {https://doi.org/10.1145/3611643.3616243},
doi = {10.1145/3611643.3616243},
booktitle = {Proceedings of the 31st ACM Joint European Software Engineering Conference and Symposium on the Foundations of Software Engineering}
}

@inproceedings{
wang2024legoprover,
title={{LEGO}-Prover: Neural Theorem Proving with Growing Libraries},
author={Haiming Wang and Huajian Xin and Chuanyang Zheng and Zhengying Liu and Qingxing Cao and Yinya Huang and Jing Xiong and Han Shi and Enze Xie and Jian Yin and Zhenguo Li and Xiaodan Liang},
booktitle={The Twelfth International Conference on Learning Representations},
year={2024},
url={https://openreview.net/forum?id=3f5PALef5B}
}

@inproceedings{bayazit2025,
author = {Bayaz\i{}t, Bar\i{}\c{s} and Li, Yao and Si, Xujie},
title = {A Case Study on the Effectiveness of LLMs in Verification with Proof Assistants},
year = {2025},
isbn = {9798400721489},
publisher = {Association for Computing Machinery},
url = {https://doi.org/10.1145/3759425.3763391},
doi = {10.1145/3759425.3763391},
booktitle = {Proceedings of the 1st ACM SIGPLAN International Workshop on Language Models and Programming Languages},
doi = {10.1145/3759425.3763391}
}

@software{the_coq_development_team_2024_11551307,
  author       = {The Coq Development Team},
  title        = {The Coq Proof Assistant},
  month        = jun,
  year         = 2024,
  publisher    = {Zenodo},
  version      = {8.19},
  doi          = {10.5281/zenodo.11551307},
  url          = {https://doi.org/10.5281/zenodo.11551307},
}

@inproceedings{mouralean2021,
author = {Moura, Leonardo de and Ullrich, Sebastian},
title = {The Lean 4 Theorem Prover and Programming Language},
year = {2021},
isbn = {978-3-030-79875-8},
publisher = {Springer-Verlag},
doi = {10.1007/978-3-030-79876-5_37},
booktitle = {Automated Deduction – CADE 28: 28th International Conference on Automated Deduction, Virtual Event, July 12–15, 2021, Proceedings}
}

@article{kumar2025saarthi,
  title={Saarthi: The first ai formal verification engineer},
  author={Kumar, Aman and Gadde, Deepak Narayan and Radhakrishna, Keerthan Kopparam and Lettnin, Djones},
  journal={arXiv preprint arXiv:2502.16662},
  year={2025}
}

@article{pothireddypalli2026agentic,
  title={Agentic AI-based Coverage Closure for Formal Verification},
  author={Pothireddypalli, Sivaram and Raman, Ashish and Gadde, Deepak Narayan and Kumar, Aman},
  journal={arXiv preprint arXiv:2603.03147},
  year={2026}
}

@article{tu2025agentic,
  title={Agentic Program Verification},
  author={Tu, Haoxin and Zhao, Huan and Song, Yahui and Zafar, Mehtab and Meng, Ruijie and Roychoudhury, Abhik},
  journal={arXiv preprint arXiv:2511.17330},
  year={2025}
}

\appendix

\end{document}